\documentclass[aps,prb,twocolumn,showpacs,floatfix,twoside,superscriptaddress,eqsecnum]{revtex4-1}
\usepackage{amssymb}
\usepackage{amsmath}
\usepackage{graphicx}
\usepackage[utf8]{inputenc}
\usepackage[colorlinks=true,allcolors=blue,dvipdfm=true]{hyperref}

\begin{document}

\title{Charge and current orders in the spin-fermion model with overlapping
hot spots.}
\author{Pavel A. Volkov}
\affiliation{Theoretische Physik III, Ruhr-Universit\"{a}t Bochum, D-44780 Bochum, Germany}
\author{Konstantin B. Efetov}
\affiliation{Theoretische Physik III, Ruhr-Universit\"{a}t Bochum, D-44780 Bochum, Germany}
\affiliation{National University of Science and Technology \textquotedblleft
MISiS\textquotedblright , Moscow, 119049, Russia}
\affiliation{International Institute of Physics, UFRN, 59078-400 Natal, Brazil}
\date{\today }

\begin{abstract}
Experiments carried over the last years on the underdoped cuprates have revealed a variety of symmetry-breaking phenomena in the pseudogap state.
Charge-density waves, breaking of $C_{4}$ rotational symmetry as well as time-reversal symmetry breaking have all been observed in several cuprate
families. In this regard, theoretical models where multiple
non-superconducting orders emerge are of particular interest. We consider
the recently introduced (Phys. Rev. B \textbf{93}, 085131 (2016))
spin-fermion model with overlapping 'hot spots'
on the Fermi surface. Focusing on the particle-hole instabilities we obtain
a rich phase diagram with the chemical potential relative to the dispersion at $(0,\pi);\;(\pi,0)$ and the Fermi surface
curvature in the antinodal regions being the control parameters. We find
evidence for d-wave Pomeranchuk instability, d-form factor charge density
waves as well as commensurate and incommensurate staggered bond current
phases similar to the d-density wave state. The current orders are found to
be promoted by the curvature. Considering the appropriate parameter range
for the hole-doped cuprates, we discuss the relation of our results to the
pseudogap state and incommensurate magnetic phases of the cuprates.
\end{abstract}

\pacs{aaaa}
\maketitle

\section{Introduction}

Origin of the pseudogap state\cite%
{timusk.1999,norman.2005,hashimoto.2014} remains one of the main puzzles in
the physics of the high-T$_{c}$ cuprate superconductors. First observed in NMR measurements\cite{warren.1989,alloul.1989}, it is characterized by the loss of the density of states due to the opening of a partial gap at the Fermi level below the
pseudogap temperature $T^{\ast }>T_{c}$. Studies of the pseudogap by means of ARPES\cite{damascelli.2003,hashimoto.2014} and Raman scattering
\cite{benhabib.2015,loret.2017} have revealed that it opens around $(0,\pi)$ and $(\pi,0)$ points of the 2D Brillouin zone, the so-called antinodal regions. With increasing hole doping both $T^*$ and the gap magnitude decrease monotonously and eventually disappear. However, modern experiments add
many more unconventional details to this picture, showing that a crucial role in the pseudogap state is played by the various ordering tendencies.

To begin with, the point-group symmetry of the $CuO_2$ planes appears to be
broken. Namely, scanning tunneling microscopy (STM)%
\cite{kohsaka.2007,lawler.2010} and transport\cite{daou.2010,cyr.2015}
studies show the absence of C$_4$ rotational symmetry in the pseudogap
state. More recently, magnetic torque measurements of the bulk magnetic
susceptibility\cite{sato.2017} confirmed C$_4$ breaking occurring at $T^*$.
Additionally, an inversion symmetry breaking associated with pseudogap has
been discovered by means of second harmonic optical anisotropy measurement%
\cite{zhao.2017}.

Other experiments suggest that an unconventional time-reversal symmetry
breaking is inherent to the pseudogap. Polarized neutron diffraction studies
of different cuprate families reveal a magnetic signal commensurate with the
lattice appearing below $T^*$ and interpreted as being due to a $\mathbf{Q}%
=0 $ intra-unit cell magnetic order\cite{fauque.2006,sidis.2013}. The signal
has been observed to develop above $T^*$ with a finite correlation length%
\cite{mangin.2015} and breaks the C$_4$ symmetry\cite{mangin.2017} (note,
however, that a recent report\cite{croft.2017} does not bear evidence of
such a signal). Additionally, at a temperature $T_K$ that is below $T^*$ but
shares a similar doping dependence polar Kerr effect has been observed\cite%
{xia.2008,he.2011}, which implies\cite{kapitulnik.2015,cho.2016} that
time-reversal symmetry is broken. Additional signatures of a temporally
fluctuating magnetism below $T^*$ are also available from the recent $\mu$Sr
studies\cite{zhang.2017,pal.2017}.

While the signatures described above indicate that the pseudogap state
is a distinct phase with a lower symmetry, there exist
only few experiments\cite{timusk.1999,shekhter.2013} that yield a
thermodynamic evidence for a corresponding phase transition. On the other
hand, transport measurements suggest the existence of quantum critical
points (QCPs) of the pseudogap phase\cite{badoux.2016}, accompanied by
strong mass enhancement\cite{ramshaw.2015} in line with the existence of a
QCP.

Additionally, in the recent years the presence of charge density waves (CDW)
has been discovered in a similar doping range. The CDW onset temperature can
be rather close to $T^{\ast }$\cite{comin.2014} but has been shown to have a
distinct dome-shaped doping dependence in YBCO\cite{huecker.2014} and Hg-1201%
\cite{tabis.2017}. Diverse probes such as resonant\cite%
{comin.2014,tabis.2014,comin.2016,tabis.2017} and hard X-ray\cite%
{chang.2012,huecker.2014,forgan.2015,campi.2015} scattering, STM\cite%
{wise.2008,lawler.2010,parker.2010,fujita.2014,hamidian.2016} and NMR\cite%
{wu.2011,wu.2015} have observed CDW with similar properties in most of the
hole-doped cuprate compounds with the exception of La-based ones (in which
the spin and charge modulations are intertwined\cite{tranquada.2013}).
Generally, the CDWs have the following common properties.
The modulation wavevectors are oriented along the Brillouin zone axes (axial
CDW) and decrease with doping. While the modulations along both
directions are usually observed, there is an experimental
evidence\cite{campi.2015,hamidian.2016,comin.2015} that the CDW is
unidirectional locally. The intra-unit cell structure of the CDW is
characterized by a d-form-factor\cite{comin.2015.nmat} with the charge being
modulated at two oxygen sites of the unit cell in antiphase with each other.

From the theoretical perspective, one of the initial interpretations was that the pseudogap was a manifestation of fluctuating superconductivity, either in a form of preformed Bose pairs\cite{randeria.1992,alexandrov.1994} or strong phase fluctuations\cite{emery.1995}. However, the onset temperatures of superconducting fluctuations observed
in the experiments\cite{li.2010,alloul.2010,yu.2017} are considerably %
below $T^{\ast }$ and have a distinct doping dependence. Another scenario
dating back to the seminal paper\cite{anderson.1987} attributes the
pseudogap to the strong short-range correlations due to strong on-site
repulsion\cite{phillips.2009,rice.2011}. Numerical quantum Monte Carlo
simulations\cite{gull.2010,sordi.2012,gunnarson.2015,fratino.2016} of the
Hubbard model support this idea. However, this scenario 'as is'
does not explain the broken symmetries of the pseudogap state. More
recently, these results have been interpreted as being due to topological
order\cite{wu.2017,scheurer.2017}, that can also coexist with the breaking
of discrete symmetries\cite{chatterjee.2017}.

A different class of proposals for explaining the pseudogap behavior involves a competing symmetry-breaking order. One of the possible candidates discussed in the literature is the $\mathbf{Q}=0$ orbital loop
current order\cite{varma}. Presence of circulating currents explicitly breaks the time reversal symmetry,
allowing one, with appropriate modifications, to describe the phenomena observed in polarized neutron scattering\cite{he.varma.2011} and polar Kerr effect\cite{aji.2013} experiments. However, it does not lead to a gap on the Fermi surface at the mean-field level. Numerical studies of the three-band Hubbard
model give arguments both for\cite{weber.2009,weber.2014} and against \cite{thomale.2008,nishimoto.2009,kung.2014} this type of order. Other proposals for $\mathbf{Q}=0$ magnetic order include spin-nematic\cite{fauque.2006,fischer.2014}, oxygen orbital moment\cite{moskvin.2012} or magnetoelectric multipole\cite{lovesey.2015,fechner.2016} order.

Charge nematic order\cite{kivelson.1998} and the related d-wave Pomeranchuk instability\cite{halboth.2000,yamase.2000} of the Fermi surface have also been considered in the context of the pseudogap state. It breaks the C$_4$ rotational symmetry of the $CuO_2$ planes in agreement with numerous experiments\cite{kohsaka.2007,lawler.2010,daou.2010,cyr.2015,sato.2017}. While not opening a gap, fluctuating distortion of the Fermi surface can result in an arc-like momentum distribution of the spectral weight and non-Fermi liquid behaviour\cite{delanna.2007,yamase.2012,lederer.2017}. Evidence for this order comes from numerical studies of the Hubbard model with functional renormalization group\cite{halboth.2000}, dynamical mean-field theory\cite{okamoto.2012,kitatani.2017} and other\cite{kaczmarczyk.2016,zheng.2016} methods as well as analytical studies of forward-scattering\cite{yamase.2005} and spin-fermion\cite{volkov.2016,volkov.2.2016} models.

Another possibility is the CDW\cite{castellani.1995,li.2006,gabovich.2010}.
More recent studies focus on the important role of the interplay between CDW
and superconducting fluctuations \cite{efetov.2013}, preemptive orders and time reversal symmetry breaking
\cite{wang.2014} (that can result in the polar Kerr effect\cite{wang.2014.2,gradhand.2015}), vertex corrections for the interactions\cite%
{yamakawa.2015,tsuchiizu.2016}, CDW phase fluctuations\cite{caprara.2017}
and possible SU(2) symmetry\cite{pepin.2014,kloss.2016,morice.2017}.
Additionally, pair density wave - a state with modulated Cooper pairing
amplitude has been proposed to explain the pseudogap and CDW\cite{lee.2014,wang.2015,wang.2015.2},
which can be also understood with the concept of 'intertwined' SC
and CDW orders\cite{fradkin.2015}.

An interesting alternative is the d-density wave
\cite{chakravarty.2001} (DDW) state (also known as flux phase%
\cite{affleck.1989,marston.1989}) which is characterized by a pattern of
\textit{bond currents} modulated with the wavevector $\mathbf{Q}=(\pi ,\pi )$
that is not generally accompanied by a charge modulation. This order leads to
a reconstructed Fermi surface consistent with the transport\cite%
{badoux.2016,storey.2017} and ARPES\cite{hashimoto.2010} signatures of the
pseudogap. Moreover, the time-reversal symmetry is also broken and a
modified version of DDW can explain the polar Kerr effect\cite{sharma.2016}
observation. Additionally, model calculations show\cite%
{atkinson.2016,makhfudz.2016} that the system in the DDW state can be
unstable to the formation of axial CDWs. Studies aimed at a direct
detection of magnetic moments created by the DDW have yielded results both
supporting\cite{mook.2002,mook.2004} and against\cite{stock.2002} their existence
(or with the conclusion that the signal is due to impurity phases%
\cite{sonier.2009}). However, theoretical estimates of the
resulting moments are model-dependent\cite{hsu.1991,chakravarty.kee.2001}.
There also exists indirect evidence from superfluid density measurements\cite%
{trunin.2004}. Theoretical support for the DDW comes from renormalization
group\cite{honerkamp.2002} and variational Monte Carlo\cite{yokoyama.2016}%
studies of the Hubbard model, DMRG studies of $t-J$ ladders\cite%
{schollwock.2003}, and mean-field studies of $t-J$\cite{raczkowski.2007} as
well as single\cite{laughlin.2014} and three-orbital\cite{bulut.2015}
models. However, the regions of DDW stability found in these studies vary significantly
and depend on the value of
particular interactions\cite{laughlin.2014,bulut.2015} or details of the
Fermi surface \cite{yokoyama.2016}.

Overall, the question of possible competing orders in the cuprates has
turned out to be a rather complicated one. Interestingly, state-of-art
numerical calculations comparing different methods show that the energy
difference between distinct ground states can be miniscule\cite{zheng.2017}
explaining some of the difficulties. Thus, analytical approaches which allow one
to study the influence of different parameters in detail can be of interest.

In this paper, we deduce leading non-superconducting orders using a low-energy effective theory for fermions interacting with antiferromagnetic (AF) fluctuations. While such theories can be in principle derived from the microscopic Hubbard or t-J Hamiltonians\cite{kochetov.2015}, we employ here a semiphenomenological approach in the spirit of the widely used spin-fermion (SF) model\cite{abanov.2003,metlitski.2010,efetov.2013,wang.2014,pepin.2014}. Our take on this problem differs in that we relax the usual assumption that the interaction, being peaked at $(\pi,\pi)$, singles out eight {\it isolated} 'hot spots' on the Fermi surface. In contrast, we consider that neighboring hot spots may strongly overlap and form antinodal 'hot regions'. This assumption agrees well with the ARPES results\cite{hashimoto.2014} demonstrating the pseudogap covers the full antinodal region without pronounced maxima at the 'hot spots' of the standard SF model. Moreover, the electron spectrum in the antinodal regions has been found\cite{hashimoto.2010,kaminski.2006} to be shallow with respect to the pseudogap energy scale for the hole-underdoped samples, i.e. the pseudogap opens also at points that are not in immediate vicinity of the Fermi surface. From the spin fluctuation perspective this can be anticipated if the AF fluctuations correlation length is small enough such that the resulting interaction between fermions is uniformly smeared covering the full antinodal regions. Indeed, the neutron scattering experiments\cite{haug.2010,chan.2016} show that the correlation lengths at the temperatures and dopings relevant for the pseudogap amount to several unit cells lengths.

SF model with overlapping hot spots has been introduced in our recent
publications\cite{volkov.2016,volkov.2.2016}, where we have
considered normal state properties as well as charge orders corresponding
to intra-region particle-hole pairing. For the case of a small
Fermi surface curvature, it has been shown that the d-wave Pomeranchuk instability is the leading one for sufficiently shallow electron spectrum in the hot regions. This is in contrast to the diagonal d-form factor CDW usually being the leading particle-hole instability in the standard SF model \cite{metlitski.2010,efetov.2013,pepin.2014}. As a result of Pomeranchuk transition, the $C_{4}$ symmetry gets broken by a
deformation of the Fermi surface and an intra-unit-cell charge redistribution.
Additionally, as the Pomeranchuk order leaves the Fermi surface ungapped, we
have shown that at lower temperatures an axial CDW with dominant d-form
factor and d-wave superconductivity may appear. These results are
in line with the simultaneous observation of the commensurate C$%
_{4} $ breaking\cite{kohsaka.2007,lawler.2010,daou.2010,cyr.2015,sato.2017}
and axial d-form factor CDWs\cite{fujita.2014,comin.2015.nmat}. At
the same time, these order parameters, although being in agreement
with the experimental observations, do not readily explain the
time-reversal symmetry breaking phenomena as well as the possible Fermi
surface reconstruction into hole pockets\cite{badoux.2016}.

In this paper, we consider a possibility of an inter-region particle-hole pairing, akin to the excitonic insulator proposed
long ago\cite{KK}. The resulting state is similar in properties to the
d-density wave\cite{chakravarty.2001}, having staggered bond currents.
In addition, we find also evidence for an
incommensurate version thereof. It turns out (Sec. \ref{sec:tm},\ref{sec:sf}%
) that the Fermi surface curvature in the antinodal regions (assumed to be
small in Refs.\onlinecite{volkov.2016,volkov.2.2016}) is the most
important ingredient that stabilizes this state against the charge orders,
thus leading to a rich phase diagram. We further discuss the
relation of our findings to the pseudogap state in Sec.\ref{sec:disc}.

The paper is organized as follows. In Sec. \ref{sec:model} we present the
model and assumptions that we use. In Sec. \ref{sec:tm} we analyze a
simplified version of the model ignoring retardation effects and identify
the emerging orders. In Sec. \ref{sec:sf} we present the results for the
full model and discuss the approximations used. In Sec. \ref{sec:disc} we
discuss the relation of our results to the physics of underdoped cuprates
and in Sec.\ref{sec:concl} we summarize our findings.

\section{Model.}
\label{sec:model}
\begin{figure}[h]
\includegraphics[width=\linewidth]{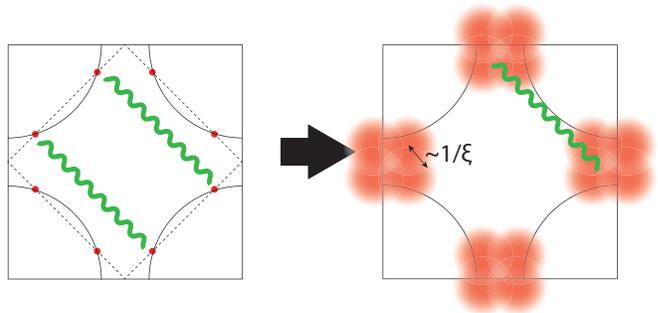}
\caption{Overlapping hot spots on the Fermi surface typical for the hole-doped cuprates.}
\label{Fig:sfoverlap}
\end{figure}

The spin-fermion model describes the low-energy
physics of the cuprates in terms of low-energy fermions
interacting via the antiferromagnetic paramagnons. The
latter are assumed to be remnants of the parent insulating
AF state destroyed by hole doping\cite{abanov.2003}. The resulting interaction is strongly peaked at the
wavevector $\mathbf{Q}_{0}=(\pi ,\pi )$ corresponding to the
antiferromagnetic order periodicity and is described by a propagator
\begin{equation*}
\frac{1}{(\mathbf{q}-\mathbf{Q}_{0})^{2}+\frac{1}{\xi _{AF}^{2}}-\left(
\frac{\omega }{v_{s}}\right) ^{2}}.
\end{equation*}%
Then, one can identify eight 'hot spots' on the Fermi Surface
mutually connected by $\mathbf{Q_{0}}$ where the interaction is
expected to be strongest (see the left part of Fig.\ref{Fig:sfoverlap}). A
conventional approximation motivated by the proximity to AF quantum critical
point (QCP) where $\xi _{AF}\rightarrow \infty $ is to consider only small $%
\delta \mathbf{p}\sim 1/\xi _{AF}$ vicinities of the 'hot spots' to be
strongly affected by the interaction. However, at temperatures relevant for
the pseudogap state this argument does not have to hold --- the
experimentally reported correlation lengths\cite{haug.2010,chan.2016} are
indeed rather small. Moreover, ARPES experiments\cite{hashimoto.2014} show
that the effects of the pseudogap extend well beyond the 'hot spots' to the
Brillouin zone edges $(\pi ,0),(0,\pi )$ without being significantly
weakened.

A different approach has been introduced in \cite{volkov.2016,volkov.2.2016}
. As $\xi _{AF}$ becomes smaller, the 'hot spots' expand and and can
eventually overlap and merge forming two 'hot regions' (see the right part
of Fig.\ref{Fig:sfoverlap}). For the latter to occur the fermionic
dispersion in the antinodal region should be shallow (see also the
formal definition below), which is supported by the experimental
data\cite{hashimoto.2010,kaminski.2006}. To describe this situation we
consider the following Lagrangian:
\begin{equation}
\begin{gathered} L_{\mathrm{SF}} =\sum_{\mathbf{p},\nu =1,2} c_{\bf p}^{\nu
\dagger } \left[\partial _{\tau }+\varepsilon _{\nu}({\bf p}) \right] c
_{\bf p}^{\nu } \\ +\frac{1}{2}\sum_{\bf q} \vec{\varphi}_{-{\bf q}}
(-v_{s}^{-2}\partial _{\tau }^{2}+{\bf q}^2+1/\xi^2) \vec{\varphi}_{\bf q}
\\ +\lambda \sum_{\mathbf{p,q}} \left[ c _{\mathbf{p+q}}^{1\dagger }
\vec{\varphi}_{\mathbf{q}}\vec{\sigma}c _{\mathbf{p}}^{2} +c
_{\mathbf{p+q}}^{2\dagger}\vec{\varphi}_{\mathbf{q}}\vec{\sigma}c_{%
\mathbf{p}}^{1} \right], \end{gathered}  \label{Mod:lagr}
\end{equation}%
where $c_{\mathbf{p}}^{\nu }$ and $\vec{\varphi}_{\mathbf{q}}$ are the
fermionic and bosonic (paramagnon) fields, respectively and $\varepsilon
_{\nu }(\mathbf{p})$ is the fermionic dispersion where the index $\nu $
enumerates the two 'hot regions'. Additionally, as the quantity $%
|\varepsilon ((0,\pi )/(\pi ,0))-\mu |$ has been observed\cite%
{hashimoto.2014,hashimoto.2010} to be of the order of the pseudogap energy
or smaller, one expects the fermions in the whole region to participate in
the interaction. Consequently one has to consider the dispersion relation
not linearized near the Fermi surface. Due to the saddle points present at $%
(\pi ,0)$ and $(0,\pi )$ the minimal model for the dispersion is:
\begin{equation}
\varepsilon _{1}(\mathbf{p})=\alpha p_{x}^{2}-\beta p_{y}^{2}-\mu
,\;\varepsilon _{2}(\mathbf{p})=\alpha p_{y}^{2}-\beta p_{x}^{2}-\mu ,
\label{Mod:disp}
\end{equation}%
where $\alpha $ has the meaning of the inverse fermion mass and $\beta
/\alpha $ controls the Fermi surface curvature. Note that the
chemical potential of the system is determined by the full Fermi surface. As
the Fermi energies (measured from the $\Gamma $-point) in hole-doped
cuprates are quite large we neglect the temperature dependence of the
chemical potential, and, consequently, $\mu$. On the other hand, as the relevant temperatures for the pseudogap onset are still sizable, we shall not consider the effects of simultaneous development of multiple instability channels due to the van Hove singularities\cite{lehur.2009}.

In Refs. \onlinecite{volkov.2016,volkov.2.2016} the limit $\beta \rightarrow 0$ has
been considered. For this case the condition of 'shallowness' leading to the
merging of the hot spots reads $1/\xi _{AF}\gg \sqrt{\mu /\alpha }$%
. However, in order to keep the simple quadratic form of the
paramagnon dispersion we also assume $\xi _{AF}\gg a_{0}$, where $a_{0}$ is
the lattice spacing. Here we will consider the consequences of finite $\beta
$ for the model \ref{Mod:lagr} under the same assumption of the strong hot
spot overlap. As in Refs. \onlinecite{volkov.2016,volkov.2.2016} we will
concentrate on the particle-hole (non-superconducting) orders. Note that for interaction being via the antiferromagnetic paramagnons only, d-wave superconductivity is expected to overcome the particle-hole orders\cite{metlitski.2010,efetov.2013}. However, additional interactions present in real systems, such as nearest-neighbor\cite{sau.2014} or remnant low-energy Coulomb repulsion\cite{volkov.2016} should suppress it with respect to the particle-hole orders.

\section{Phase Diagram for a Simplified Model.}

\label{sec:tm} To address qualitative features of the emerging
orders we can use a simplified version of the model (\ref{Mod:lagr})
also introduced previously by us\cite{volkov.2016}. It amounts to
substitution of the paramagnon part of the Lagrangian with a constant
interregion interaction between the fermions. This is also equivalent to
taking the $\xi \rightarrow 0$ limit for the paramagnon propagator.
Additionally, we neglect the self-energy effects for this case.

We start with the following Lagrangian
\begin{equation}
\begin{gathered} \sum_{{\bf p},\nu =1,2,\sigma} c_{{\bf p}}^{\nu \dagger } [
\partial _{\tau }+\varepsilon_i({\bf p})] c_{{\bf p}}^{\nu } \\
-\frac{\lambda_0}{6}
\sum_{{\bf p},{\bf p}',{\bf q}}
 [ c_{\bf p+q}^{1 \dagger } \vec{\sigma} c_{\bf p}^2+
c_{\bf p+q}^{2 \dagger } \vec{\sigma} c_{\bf p}^1 ] 
\\
\times[ c_{\bf p'-q}^{1
\dagger } \vec{\sigma} c_{\bf p'}^2+ c_{\bf p'-q}^{2 \dagger } \vec{\sigma}
c_{\bf p'}^1 ]. \end{gathered}  \label{tm:lagr}
\end{equation}%
The spin structure of the interaction is taken here in full
analogy to the original spin-fermion model. Additionally, the
integrals that appear below are cut off at momenta $p_{x},\;p_{y}\sim 1/\xi $
and we assume that the inequality $\alpha /\xi ^{2}\gg \mu ,T$
(strong overlap of hot spots) is fulfilled.

\begin{figure}[h]
\includegraphics[width=\linewidth]{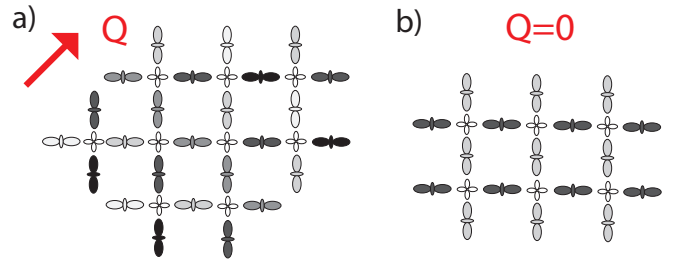}
\caption{Illustration of charge modulations for a) d-form factor CDW with $%
\mathbf{Q}$ along the diagonal b) d-wave Pomeranchuk phase}
\label{Fig:cdwpic}
\end{figure}

In addition to the d-wave superconductivity, one can
identify two attractive singlet channels (the triplet channels, as in the SF
model, are subleading with the effective coupling being three times weaker).
Corresponding order parameters can be written in terms of the averages (spin
indices are suppressed):
\begin{align}
W& =\langle c_{\mathbf{p+Q}/2}^{1\dagger }c_{\mathbf{p-Q}/2}^{1}\rangle
=-\langle c_{\mathbf{p+Q}/2}^{2\dagger }c_{\mathbf{p-Q}/2}^{2}\rangle ,
\label{tm:chargeord} \\
D& =\langle c_{\mathbf{p+Q}/2}^{1\dagger }c_{\mathbf{p-Q}/2}^{2}\rangle
=-\langle c_{\mathbf{p+Q}/2}^{2\dagger }c_{\mathbf{p-Q}/2}^{1}\rangle .
\label{tm:curord}
\end{align}%
The interaction for the orders without the sign change between the regions
(corresponding to s-form factor charge order) is repulsive and consequently
such orders are not expected to appear on the mean-field level. As is shown
in detail below, the order parameter $W$, Eq.(\ref%
{tm:chargeord}), corresponds to a charge order with a d-form
factor. Due to the change of the sign between the regions the
charge is modulated only at the oxygen orbitals of the unit cell
corresponding to the bonds in a single-band model. In the case $\mathbf{Q}%
\neq 0$ (Fig.\ref{Fig:cdwpic}a) this order represents the d-form factor
charge density wave while for $\mathbf{Q}=0$ (Fig.\ref{Fig:cdwpic}b) it
leads to an intra-unit cell redistribution of charge accompanied by a d-wave
Pomeranchuk deformation of the Fermi surface.

\begin{figure}[h]
\includegraphics[width=\linewidth]{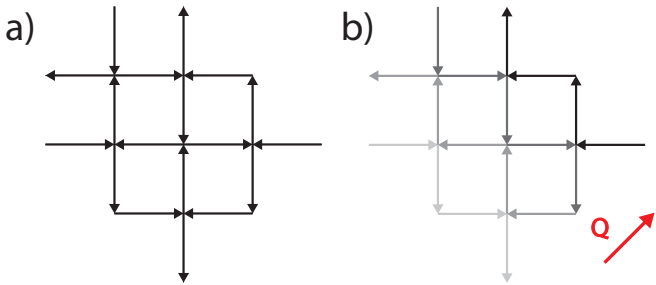}
\caption{Illustration of current modulations for a) DDW b) IDDW with $%
\mathbf{Q}$ along the diagonal. Only the currents between nearest neighbors
are depicted.}
\label{Fig:ddwpic}
\end{figure}

A non-zero average $D$,\ Eq. (\ref{tm:curord}), on the other hand,
leads to a pattern of bond currents modulated with wavevector $(\pi ,\pi )+
\mathbf{Q}$ without any charge density modulation. In the case $
\mathbf{Q}=0$ (see Fig.\ref{Fig:ddwpic}a) this state is similar to the DDW%
\cite{chakravarty.2001} or the staggered flux phase\cite{marston.1989}.
Additionally, the order parameter is purely imaginary in this case ($D=-D^{\ast }$) and therefore breaks only a discrete symmetry. Finite $\mathbf{Q}\neq 0$ correspond to a modulation of
the current pattern incommensurate with the lattice (see Fig.\ref{Fig:ddwpic}%
b), which results in a breaking of a continuous rather then
discrete symmetry. We will call the resulting state incommensurate DDW
(IDDW).

In order to obtain a phase diagram containing the orders discussed above in for the model described by the Lagrangian (\ref{tm:lagr})
we calculate critical temperatures of the corresponding
transitions using linearized mean field equations
\begin{align}
\frac{1}{\nu _{0}\lambda _{0}}& =\chi _{W}^{1}(T_{W},\mathbf{Q})=\chi
_{W}^{2}(T_{W},\mathbf{Q}),  \label{tm:tcw} \\
\frac{1}{\nu _{0}\lambda _{0}}& =\chi _{D}(T_{D},\mathbf{Q}),  \label{tm:tcd}
\end{align}%
where $\nu_0=S/(2\pi)^2$, $S$ being the area of the 2D system and
\begin{equation}
\begin{gathered} \chi_{W}^l(T,{\bf Q})=T \sum_{n}\int d{\bf p} \frac{-1}{(i
\omega_n- \varepsilon^l_{{\bf p}+{\bf Q}/2}) (i \omega_n-
\varepsilon^l_{{\bf p}-{\bf Q}/2})}, \\ \chi_{D}(T,{\bf Q})=T\sum_{n}\int
d{\bf p} \frac{-1}{(i \omega_n- \varepsilon^1_{{\bf p}+{\bf Q}/2})(i
\omega_n- \varepsilon^2_{{\bf p}-{\bf Q}/2})}, \end{gathered}  \label{tm:chi}
\end{equation}%
where $\omega _{n}=(2n+1)\pi T$ is the fermionic Matsubara frequency.
Evaluating the susceptibilities (\ref{tm:chi}) one can find the
critical temperatures for the instabilities considered here. To map out the phase diagram we fix the leading instability temperature $T_{ins}$ and identify the order having the largest $\chi $ as the leading
one. The control parameters are then $\mu /T_{\mathrm{ins}}$ and $\beta
/\alpha $.

\begin{figure}[h]
\includegraphics[width=\linewidth]{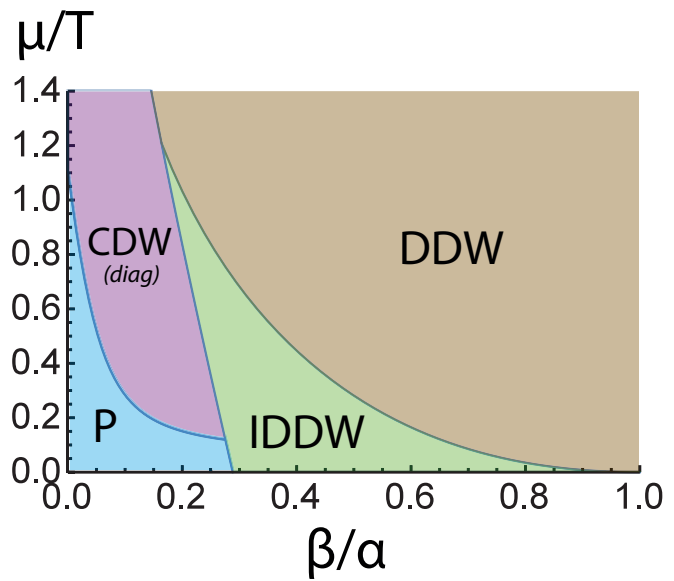}
\caption{Phase diagram of the model (\protect\ref{tm:lagr}). 'P' is for the
d-wave Pomeranchuk instability, 'CDW(diag)' - is the d-form factor charge
density wave with wavevector along the BZ diagonal, 'DDW' ('IDDW') stand for
(incommensurate) d-density wave ($\mathbf{Q}\neq0$ in (\protect\ref%
{tm:curord})). Axial CDW can emerge from Pomeranchuk or DDW phases at a
lower temperature (see text).}
\label{Fig:tmphasediagr}
\end{figure}

In Fig. \ref{Fig:tmphasediagr} we present the resulting phase diagram where $%
\chi _{W,D}$ have been calculated using $\alpha /\left( \xi
^{2}T\right)=50$. In contrast to the previous studies of
the SF model\cite{metlitski.2010,efetov.2013,sau.2014}, where a diagonal
d-form factor CDW has been universally found to be the leading particle-hole order,
one obtains now three novel instabilities here: to the d-wave
Pomeranchuk phase with a deformed Fermi surface and intra-unit cell charge
nematicity, as well as current orders in the form of DDW and its
incommensurate variation (IDDW). Note that the experimentally observed\cite%
{fujita.2014,comin.2015.nmat} axial d-form factor CDW is a
subleading instability, which is however expected to occur at lower
temperatures within the Pomeranchuk\cite{volkov.2016} or DDW\cite{atkinson.2016} phases.

Two general trends are evident from Fig. \ref{Fig:tmphasediagr}. First,
the charge order is generally favorable at low
Fermi surface curvature $\beta /\alpha $ while the current orders
dominate at larger ones. Secondly, small values of $\mu $ are seen
to stimulate the Pomeranchuk and IDDW phases among the charge and current
phases, respectively.

The qualitative reason for the dominance of the DDW at moderate $\beta
/\alpha $ can be seen in the dispersion (\ref{Mod:disp}) : for $\beta
=\alpha $ one has $\varepsilon _{1}=-\varepsilon _{2}$ for $\alpha p^{2}\gg
\mu $. As $\alpha /\xi ^{2}\gg \mu $, the Fermi surfaces of the two regions
are nearly nested in the large part of the regions, in contrast to a CDW,
for which nesting is restricted to a vicinity of a single point in $\mathbf{p}$-space. It is, however, surprising that the current phases
start dominating at $\beta /\alpha $ considerably smaller than 1.
Below we provide analytical results leading to the phase diagram of Fig. \ref{Fig:tmphasediagr} as well as a detailed description of the emerging orders.

$\bullet $ \textit{Charge Density Wave} is represented by Eq. (\ref%
{tm:chargeord}) with finite value of $\mathbf{Q}$. Due to the sign
change between the two FS regions the amplitude of the on-site charge
modulation proportional to $\sum_{\mathbf{p}}W(\mathbf{p})\sim
W_{1}+W_{2}$ vanishes.

It has been shown\cite{efetov.2013,volkov.2016}, however, that the charge
modulation on oxygen sites is related to bond operators in the single-band
model as:
\begin{equation*}
\delta n_{O}\sim \delta \langle c_{i+1,\sigma }^{\dagger }c_{i,\sigma
}+c_{i,\sigma }^{\dagger }c_{i+1,\sigma }\rangle _{CO},
\end{equation*}%
where $i$ and $i+1$ are two neighboring copper sites. The proportionality
coefficient depends on additional assumptions: in Ref. %
\onlinecite{volkov.2016} $p/8$ has been obtained using Zhang-Rice singlet
doping picture. Transforming the expression to the momentum space one
obtains
\begin{equation}
\begin{gathered} \delta n_{O_{x}}(\mathbf{r}) \sim e^{i\mathbf{Q}\mathbf{r}}
(W_1-W_2)+c.c.,\\ \delta n_{O_{y}}(\mathbf{r}) \sim
-e^{i\mathbf{Q}\mathbf{r}}(W_1-W_2)+c.c. \label{tm:ox} \end{gathered}
\end{equation}

Let us compute now the corresponding susceptibility, Eq. (%
\ref{tm:chi}). First of all, one can conclude from Eq. (\ref{tm:tcw}%
) that only $Q_{x}=Q_{y}\equiv Q$ satisfies $\chi _{W}^{1}=\chi _{W}^{2}$,
i.e. the wavevector is directed along the diagonal. This is in line with the
previous results on the spin-fermion model\cite{metlitski.2010,efetov.2013}.
The momentum integrals for the present model with overlapping
hotspots can be evaluated explicitly for two limiting cases (for the
details of calculation see Appendix \ref{app:tm}). In the limit $%
\beta \rightarrow 0$ one obtains:
\begin{equation}
\chi _{CDW}(T,\mathbf{Q})=\frac{T}{\alpha }\sum_{\omega _{n}}\frac{i\pi \;%
\mathrm{sgn}[\omega _{n}]\sqrt{\alpha /\xi ^{2}}}{(i\omega _{n}+\mu -\alpha
Q^{2}/4)\sqrt{i\omega _{n}+\mu }}.  \label{tm:chicdw0}
\end{equation}%
In the opposite limit $\beta /\xi ^{2}\gg \mu ,T$ one gets
\begin{equation}
\chi _{CDW}(\mathbf{Q})\approx \frac{T}{\sqrt{\alpha \beta }}\sum_{|\omega
_{n}|<\beta /\xi ^{2}}\frac{-i\pi \mathrm{sgn}[\omega _{n}]\mathrm{arctanh}%
\sqrt{\frac{\gamma }{\gamma -i\omega _{n}-\mu }}}{\sqrt{\gamma }\sqrt{\gamma
-i\omega _{n}-\mu }},  \label{tm:chicdw1}
\end{equation}%
where $\gamma =(\alpha -\beta )Q^{2}/4$. The expression in the sum in Eq. (\ref{tm:chicdw1}) is obtained for $\omega _{n}\ll \beta /\xi ^{2}$.
However, as the resulting sum is logarithmically large for $\beta /\xi
^{2}\gg \mu ,T$,we can simply disregard the contribution from higher Matsubara frequencies.

$\bullet $ \textit{Pomeranchuk instability} corresponds to the anomalous
average \ref{tm:chargeord} with $\mathbf{Q}=0$. It leads to a d-wave-like
deformation of the Fermi surface breaking $C_{4}$ symmetry without opening a
gap. Additionally, the Pomeranchuk order should be accompanied by an
intra-unit cell redistribution of the charge on the two oxygen
orbitals, which can be readily seen from (\ref{tm:ox}).

The expressions for the susceptibilities can be obtained from
Eqs. (\ref{tm:chicdw0}) and (\ref{tm:chicdw1}) by taking the $%
\mathbf{Q}\rightarrow 0$ limit. Moreover, the sign of $\left. \frac{\partial
\chi _{CDW}}{\partial Q^{2}}\right\vert _{\mathbf{Q}=0}$ allows one
to check the stability of the Pomeranchuk phase with respect to the CDW.

For $\beta \rightarrow 0$ we get
\begin{equation}
\chi _{Pom}=\frac{T}{\alpha }\sum_{\omega _{n}}\frac{i\pi \;\mathrm{sgn}%
[\omega _{n}]\sqrt{\alpha /\xi ^{2}}}{(i\omega _{n}+\mu )^{3/2}}.
\label{tm:chipom0}
\end{equation}%
and
\begin{equation}
\left. \frac{\partial \chi _{CDW}}{\partial Q^{2}}\right\vert _{\mathbf{Q}%
=0}=\frac{T}{4}\sum_{\omega _{n}}\frac{i\pi \;\mathrm{sgn}[\omega _{n}]\sqrt{%
\alpha /\xi ^{2}}}{(i\omega _{n}+\mu )^{5/2}}.  \label{tm:chicdwexp0}
\end{equation}%
Numerical calculation shows that the expression (\ref{tm:chicdwexp0})
changes sign from positive to negative for $\mu /T<1.1$. Therefore,
the Pomeranchuk phase is stable for $\mu /T_{ins}<1.1$ for $\beta
\rightarrow 0$ (in agreement with Ref. \onlinecite{volkov.2016}). In the
opposite limit $\beta /\xi ^{2}\rightarrow \infty $, however, the expansion
of (\ref{tm:chicdw1}) yields
\begin{equation}
\begin{gathered} \frac{\partial \chi_{CDW}}{\partial Q^2}\approx
\frac{\alpha-\beta}{4} \frac{2T}{3\alpha} \sum_{|\omega_n|<\beta/\xi^2}
\frac{i\pi{\rm sgn} [\omega_n]} {(i\omega_n+\mu)^2} \\ =
\frac{\alpha-\beta}{4} \frac{2T}{3\alpha} \sum_{0<\omega_n<\beta/\xi^2}
\frac{4\pi\omega_n\mu} {(\omega_n^2+\mu^2)^2} \end{gathered}
\end{equation}%
which is always positive. Thus, to obtain the phase boundary
between CDW and Pomeranchuk phases at finite $\beta $ one needs to perform
the momentum integration assuming finite values of $\beta /\xi ^{2}$. The general result is rather cumbersome and is presented in Appendix \ref{app:tm} (Eq. \ref{app:pommucr}). For $\mu /T\ll 1$ a simple expression is
found
\begin{equation}
\left( \frac{\mu }{T}\right) _{cr}^{CDW}=\frac{\pi ^{2}}{7\zeta (3)}\frac{%
\alpha }{\alpha -\beta }\frac{T}{\beta /\xi ^{2}}\approx 1.17\frac{\alpha }{%
\alpha -\beta }\frac{T}{\beta /\xi ^{2}}.  \label{tm:pommucr}
\end{equation}%
One can see that $(\mu /T)_{cr}$ decreases for $\beta <0.5$ and then starts
to increase. However, as we shall see, this upturn is located in the region
where DDW is the leading instability. Note that in Fig. \ref%
{Fig:tmphasediagr} the Pomeranchuk/CDW boundary is found from the full
expression (\ref{app:pommucr}) numerically for $(\alpha /\xi ^{2})/T=50$.

$\bullet $ \textit{Current Phases (DDW/IDDW)} are represented by the
anomalous average (\ref{tm:curord}). This order parameter does not
result in any charge modulations on both the copper and the oxygen orbitals. In
the former case this is guaranteed by the d-wave symmetry, while for the oxygen
orbitals (e.g., $O_{x}$):
\begin{gather*}
\delta n_{O_{x}}(\mathbf{r})\sim \\
\mathrm{Re}\left[ e^{i(\mathbf{Q}+(\pi ,\pi ))\mathbf{r}}\sum_{\mathbf{p}%
}\cos p_{x}\langle c_{\mathbf{p}+[(\pi ,\pi )+\mathbf{Q}]/2}^{\dagger }c_{%
\mathbf{p}-[(\pi ,\pi )-\mathbf{Q}]/2}\rangle \right] \\
=2\mathrm{Re}\left[ e^{i(\mathbf{Q}+(\pi ,\pi ))\mathbf{r}}\sum_{\mathbf{p}%
}\sin (p_{x})D\right] =0.
\end{gather*}%
However, it can be shown that the order parameter $D$ induces a staggered pattern of currents flowing through the lattice. The
current between the lattice sites $i$ and $j$ is given by\cite{bulut.2015}:
\begin{equation}
I_{ij}=-i\frac{t_{ij}}{\hbar }\langle \hat{c}_{i}^{\dagger }\hat{c}_{j}-\hat{%
c}_{j}^{\dagger }\hat{c}_{i}\rangle ,
\end{equation}%
where $t_{ij}$ is the hopping parameter. For example, the current $I_{i,i+x}$ between nearest neighbor sites along $x$ can be estimated as
\begin{equation}
I_{i,i+x}\approx 4\mathrm{Re}\left[ \frac{it}{\hbar }e^{i(\mathbf{Q}+(\pi
,\pi ))\mathbf{R}_{i}-iQ_{x}/2}D\right] .
\end{equation}%
In Fig. \ref{Fig:ddwpic} an illustration of the current patterns is
presented. Note that, in general, currents between non-nearest
neighbors are induced, too. However, as this effect depends on the
structure of the DDW amplitude in the entire Brillouin zone, we
will not consider it in this work.

To calculate the resulting magnetic fields one should however calculate the
\textit{current density }$\mathbf{j}_{\mathbf{q}}$ rather then the
current. For a square lattice with nearest-neighbor hopping $t$ one can
obtain the following result neglecting the smearing of atomic wavefunctions
with respect to the current variation length (see also\cite{hsu.1991,chakravarty.kee.2001}):
\begin{equation}
\begin{gathered} \mathbf{j}_{x,y}\left( \mathbf{q}\right)
=-8et\overline{D}ie^{i(\pi,0)\mathbf{e}_{x,y}}\mathbf{e}_{x,y} \\
\times \left( \mathbf{qe}_{x,y}\right) ^{-1}\sum_{\mathbf{K}_{n}}\delta
\left( \mathbf{q-Q}_{AF}+\mathbf{K}_{n}\right) , \\
\overline{D}=\frac{i}{(2\pi\xi)^2}( \langle \hat{c}_1 \hat{c}^\dagger_2 \rangle
-\langle \hat{c}_2 \hat{c}^\dagger_1 \rangle), \\
\end{gathered}
\end{equation}%
where ${\bf e}_{x(y)}$ is a unit vector along $x(y)$ axis and $\mathbf{K}_{n}$ is a reciprocal lattice vector. Additionally, one can calculate the magnetic field along the z-direction produced by the
DDW $B_{z}$ assuming that DDWs are aligned in-phase along $z$ axis
\begin{equation}
B_{z}\left(\mathbf{q}\right) =\frac{4\pi i}{c}\frac{q_{x}j_{y}-q_{y}j_{x}}{q^{2}}.
\end{equation}
Note that in the full model (Sec.\ref{sec:sf}) the order parameter is frequency-dependent due to the retardation effects and $\overline{D}$ has to be calculated using a corresponding anomalous Green's function. Thus, $\overline{D}$ is not, in general, simply related to the magnitude of the pseudogap or $T^*$ as would be the case for the constant interaction.

Let us now present the results for the thermodynamic
susceptibilities. For the commensurate (${\bf Q}=0$)
state, one obtains in the limit $\beta \rightarrow 0$
\begin{equation}
\chi _{D}(T)=\frac{\pi ^{2}}{2\alpha }\tanh {\mu /2T}.  \label{tm:chiddw0}
\end{equation}%
In the opposite limit $\beta /\xi ^{2}\rightarrow \infty ,$ we have
instead
\begin{equation}
\chi _{D}(T)\approx \frac{4\mathrm{arctanh}(\sqrt{\beta /\alpha })T}{\alpha
+\beta }\sum_{|\omega _{n}|<\beta /\xi ^{2}}\frac{i\pi \;\mathrm{sgn}[\omega
_{n}]}{i\omega _{n}+\mu }.  \label{tm:chiddw}
\end{equation}%
For the case $\mathbf{Q}\neq 0$, let us first consider the stability of the
DDW with respect to an infinitesimal discommensuration
vector $\mathbf{Q}$. The general expression obtained from $\left. \frac{%
\partial ^{2}\chi _{D}}{\partial Q_{1}^{2}}\right\vert _{\mathbf{Q}%
=0}=\left. \frac{\partial ^{2}\chi _{D}}{\partial Q_{2}^{2}}\right\vert _{%
\mathbf{Q}=0}=0$ is presented in Appendix \ref{app:tm} (Eq. \ref{app:iddwcr}%
). Numerical solution of the resulting equation is represented by
the DDW/IDDW critical line in the phase diagram of Fig. \ref
{Fig:tmphasediagr}. Qualitatively, low values of $\mu $ favor IDDW. For $\mu
/T\ll 1,\;\alpha -\beta \ll \alpha $ the result can also be
expressed analytically
\begin{equation}
\begin{gathered} \left(\frac{\mu}{T}\right)^{DDW}_{cr} \approx 0.7
(1-\beta/\alpha)^2. \end{gathered}  \label{tm:ddwmucr}
\end{equation}%
Furthermore, we have studied the dependence of the direction and magnitude
of $\mathbf{Q}$ maximizing $\chi _{DDW}(\mathbf{Q})$ numerically. For this
purpose, expressions (\ref{app:chidnum1},\ref{app:chidnum2},\ref{app:chidnum}%
) have been used.
\begin{figure}[h]
\centering
\includegraphics[width=0.8 \linewidth]{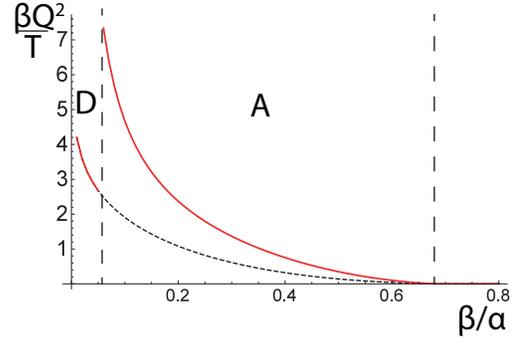}
\caption{Dependence of the IDDW incommensurability $(\protect\beta %
Q^{2}/T)_{max}$ on $\protect\beta /\protect\alpha $ for $\protect\mu /T=0.1$%
. Dashed line is $(\protect\beta Q^{2}/T)_{max}$ for $\mathbf{Q}$ along the
diagonal. Vertical lines mark transitions between different phases. In the
leftmost region (D) $\mathbf{Q}$ has diagonal orientation, in the middle (A)
- axial and in the rightmost DDW is commensurate. }
\label{Fig:tmqddw}
\end{figure}
In Fig. \ref{Fig:tmqddw} the result is presented for $\mu /T=0.1$.
Interestingly, the orientation of $\mathbf{Q}$ in the IDDW phase is
almost always along the axes. While there seems to be a transition to a
diagonal phase at low curvatures, the charge order is dominant in
that region, as is shown below.

$\bullet $ \textit{Competition between charge and current orders} We are now
in position to compare the tendencies to form charge (CDW/Pomeranchuk) and
current (DDW/IDDW) order. For $\beta \ll \alpha $, comparing (\ref{tm:chicdw0}) and (\ref{tm:chipom0}) to (\ref{tm:chiddw0}) one finds
an additional factor $\sqrt{\alpha /\xi ^{2}}$ in the former. After the
summation this translates into a large parameter of the order of $\sim \sqrt{%
(\alpha /\xi ^{2})/[T,\mu ]}$ present in the susceptibilities for
charge instabilities. Incommensurability of the DDW does not change this
conclusion (see Eq. \ref{app:chiddiag}).

In the case $\beta \sim \alpha $, $\beta /\xi ^{2}\gg \mu ,T$ one gets large
logarithmic contributions in Matsubara sums for both charge and current
orders. Therefore, one can estimate the transition line by equating the
prefactors of the sums (\ref{tm:chicdw1}) (\ref{tm:chiddw}), which
leads to the following equation
\begin{equation*}
\frac{1}{\sqrt{\alpha \beta }}=\frac{4\mathrm{arctanh}(\sqrt{\beta /\alpha })%
}{\alpha +\beta }.
\end{equation*}%
Numerical solution of this equation yields $(\beta /\alpha )_{cr}\approx
0.29 $, this value being independent of $\mu /T$. The finite slope of the
charge/current boundary in Fig. \ref{Fig:tmphasediagr} results from
finite values of $(\alpha /\xi ^{2})/T$ taken in the numerical calculations.
One can however show, that the slope is strongly suppressed being proportional to $\log ^{-1}((\beta /\xi ^{2})/T)$.

\section{Charge and Current Orders in The Full Model}

\label{sec:sf} We turn now to the analysis of the full SF model (\ref%
{Mod:lagr}). Let us first consider the effects of interactions in the normal
state. The interactions renormalize the Green's function $G$ of the fermions
and $\mathfrak{D}$ of the paramagnons leading to:

\begin{gather}
G_{\alpha \beta }^{\nu }(\varepsilon _{n},\mathbf{p})=\frac{\delta _{\alpha
\beta }}{if(\varepsilon _{n},\mathbf{p})-\varepsilon _{\nu }(\mathbf{p})},
\label{k15} \\
\mathfrak{D}_{mm^{\prime }}(\omega _{n},\mathbf{q})=-\frac{\delta
_{mm^{\prime }}}{ \Omega (\omega _{n},\mathbf{q})/v_{s}^{2}+%
\mathbf{q}^{2}+1/\xi ^{2}},  \notag
\end{gather}%
where $\alpha,\beta$ are fermion spin indices, $\nu=1,2$ is the 'hot region' index and $m,m'=(x,y,z)$ enumerate the components of the paramagnon field ${\bf \varphi}$. Additionally,$f(\varepsilon _{n},\mathbf{p})=\varepsilon
_{n}+i\Sigma (\varepsilon _{n},\mathbf{p})$ and $\Omega (\omega _{n},\mathbf{%
q})=\omega _{n}^{2}+v_{s}^{2}\Pi (\omega _{n},\mathbf{q})$, $\Sigma
(\varepsilon _{n},\mathbf{p})$ and $\Pi (\omega _{n},\mathbf{q})$ being the
fermionic self-energy and polarization operator for paramagnons,
respectively. In this section $\varepsilon _{n}=(2n+1)\pi T$ and $\omega
_{n}=2n\pi T$ stand for fermionic and bosonic Matsubara frequencies,
respectively.

\begin{figure}[h]
\includegraphics[trim = 100 150 0 0,width=0.8\linewidth]{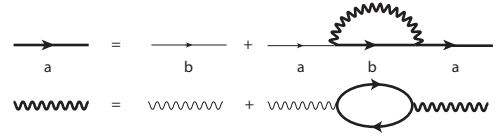}
\caption{Diagrammatic structure of approximations used, $a$ and $b$
correspond to different regions (1 or 2).}
\label{Fig:diagr}
\end{figure}

To calculate the self-energies we use the approximations illustrated
diagrammatically in Fig.\ref{Fig:diagr}. This is justified by a small
parameter $\sqrt{[T,\mu ,v_{s}^{2}/\alpha ]/(\alpha /\xi ^{2})}$ for $\beta
/\xi ^{2}\ll \mu $ (see Appendix \ref{app:sf:ver}). To study the
formation of the DDW we will, however, use these approximations for
all values of $\beta $ assuming that the results to be nevertheless
correct at least qualitatively.
The resulting momentum-dependent self-consistency equations have the same
form as the ones presented in Ref.\cite{volkov.2016}. Assuming the
strong overlap of hot spots expressed by the inequality $\alpha
/\xi ^{2}\gg \mu $, one can perform the momentum integration in the
self-energies. In the limit $\beta \ll \alpha $ the latter
can be shown to be momentum-independent. For larger $\beta $ we approximate
the self energies by their values at zero incoming momentum. Additionally, we have evaluated the momentum integral in the polarization operator for the paramagnons without any cutoff. It turns out that to reproduce our previous results\cite{volkov.2016} one needs to introduce a cutoff $\Lambda$ such that $\alpha\Lambda^2\to \infty$ while $\beta\Lambda^2\to0$. Physically, $\Lambda$ is related to the deviation of the fermionic spectrum from the form (\ref{Mod:disp}) outside the 'hot regions'. Here we will assume for simplicity that $\beta\Lambda^2\gg\mu,T$ which should be valid for not too small $\beta$ . Further details
of the calculations can be found in Appendix \ref{app:sf:eq}. Introducing an
energy scale $\Gamma =\sqrt{\lambda ^{2}v_{s}^{2}/\alpha }$ (note that in
our previous work\cite{volkov.2016} a different scale $(\lambda ^{2}v_{s}/%
\sqrt{\alpha })^{2/3}$ has been used) the resulting equations  can
be cast in a dimensionless form where all quantities are assumed to be
normalized by $\Gamma $ to the appropriate power
\begin{widetext}
\begin{equation}
\label{sf:normstate}
\begin{gathered}
f(\varepsilon)-\varepsilon =
\frac{1.5 T}{\pi}
\sum_{\varepsilon'}
\frac{
{\rm sgn}[{\rm Re} f_{\varepsilon'}]
    {\rm arctanh}
    \left\{
    \sqrt{\frac{
    \Omega_{\varepsilon-\varepsilon'}+a - (i f_{\varepsilon'}+\mu)v_s^2/\beta
    }
    {
    \Omega_{\varepsilon-\varepsilon'}+a
    }
    }
    \right\}
}
{
    \sqrt{\Omega_{\varepsilon-\varepsilon'}+a}
    \sqrt{(\Omega_{\varepsilon-\varepsilon'}+a)\beta/\alpha - (i f_{\varepsilon'}+\mu)v_s^2/\alpha}
},
\\
\Omega(\omega)-\omega^2=
\frac{T}{1+\beta/\alpha} \sum_{\varepsilon}
\frac{
{\rm sgn}[{\rm Re} f^+_{\varepsilon}]
{\rm sgn}[{\rm Re} f^-_{\varepsilon}]
}
{
\sqrt{if^+_{\varepsilon}+\mu}
\sqrt{if^-_{\varepsilon}+\mu}
}
+
\frac{4i {\rm arctanh}(\sqrt{\beta/\alpha})/\pi{\rm sgn}[{\rm Re}(f_{\varepsilon})]+1 }
{i f_{\varepsilon}+\mu}
\\
-\frac{2i}{\pi}
\frac{
   {\rm sgn}[{\rm Re} f^+_{\varepsilon}]
    {\rm arctanh}
    \left\{
    \sqrt{\frac{\alpha }{\beta}}
    \sqrt{
        \frac{i f^+_{\varepsilon} + \mu}
        {i f^-_{\varepsilon}+\mu}
    }
    \right\}
+
   {\rm sgn}[{\rm Re} f^-_{\varepsilon}]
    {\rm arctanh}
    \left\{
    \sqrt{\frac{\alpha }{\beta}}
    \sqrt{
        \frac{i f^-_{\varepsilon} + \mu}
        {i f^+_{\varepsilon}+\mu}
    }
    \right\}
}
{
\sqrt{if^+_{\varepsilon}+\mu}
\sqrt{if^-_{\varepsilon}+\mu}
},
\end{gathered}
\end{equation}
\end{widetext}
where $a=v_{s}^{2}/\xi ^{2}$,
$\Omega _{\omega }=\omega ^{2}+v_{s}^{2}\Pi (\omega )$, $f_{+}=(\alpha
f_{\varepsilon +\omega }+\beta f_{\varepsilon })/(\alpha +\beta )$ and $%
f_{-}=(\alpha f_{\varepsilon }+\beta f_{\varepsilon +\omega })/(\alpha
+\beta )$. The value $\Omega (0)$ has been absorbed into redefinition of $1/\xi ^{2}$.
Now we proceed to the analysis of the emergence of the particle-hole orders. The general mean-field equation for
the Pomeranchuk order has been derived in Ref. \cite{volkov.2016}. For the charge density wave order parameter $W_{\mathbf{Q}}(\varepsilon ,\mathbf{p})$ one obtains
\begin{gather*}
W_{\bf Q}(\varepsilon,{\bf p}) =3\lambda^2 T\sum_{\varepsilon',{\bf p}'}
\frac{\mathfrak{D}(\varepsilon-\varepsilon',{\bf p}-{\bf p}')W_{\bf Q}(\varepsilon',{\bf p}')}
{A_W},
\\
A_W=
{G_a^{-1}(\varepsilon',{\bf p}'+{\bf q}/2)G_a^{-1}(\varepsilon',{\bf p}'-{\bf q}/2)-|W_{\bf Q}(\varepsilon',{\bf p}')|^2},
\end{gather*}
while the equations for the DDW and IDDW can be written as
\begin{gather*}
D(\varepsilon,{\bf p}) = 3\lambda^2 T\sum_{\varepsilon',{\bf p}'}
\frac{\mathfrak{D}(\varepsilon-\varepsilon',{\bf p}-{\bf p}')D(\varepsilon',{\bf p}')}
{A_{D}},
\\
A_{D}=
{G_a^{-1}(\varepsilon',{\bf p}'+{\bf q}/2)G_b^{-1}(\varepsilon',{\bf p}'-{\bf q}/2)-|D_{\bf Q}(\varepsilon',{\bf p}')|^2},
\end{gather*}
These equations can be used to study the full temperature, momentum, and
Matsubara frequency dependence of the order parameters, provided the
expressions for the normal self-energies are suitably modified (see Ref.
\cite{volkov.2016} for the case of Pomeranchuk order at small $\beta $).
Here we will concentrate on the critical temperatures of the emerging orders in order to check if the general trends observed in Sec.\ref{sec:tm} still hold. Therefore, we
simply use Eqs. (\ref{sf:normstate}) for the normal state self-energies in our study.

Assuming the order parameters to be momentum independent we integrate over momenta (for details see Appendix \ref{app:sf:eq}) in the equations for charge and current orders. The resulting equations for the Pomeranchuk $%
P(\varepsilon )$, CDW $C(\varepsilon )$, DDW $D(\varepsilon )$and IDDW $%
D_{I}(\varepsilon )$ order parameters near the critical temperature are
rather cumbersome and we present them in Appendix \ref{app:sf:eq} (Eq. \ref%
{app:sf:orders}). All quantities in (\ref{app:sf:orders}) are normalized
by $\Gamma $. Additionally, to obtain a closed-form answer for IDDW, we have assumed ${\bf Q}$ along diagonal. While this assumption does not allow one to study the orientation of ${\bf Q}$, for small ${\bf Q}$ it is supposed to yield a correct critical temperature, allowing to draw conclusions about the commensurate DDW stability. From Eq. \ref{app:sf:orders} it is already evident that for $\beta /\xi ^{2}\ll 1$ r.h.s.
of the equation for the charge orders contains a large factor $1/\sqrt{%
v_{s}^{2}/\alpha }$, ultimately meaning that current orders do not appear in
this case.

\begin{figure}[h!]
\centering
\includegraphics[width=\linewidth]{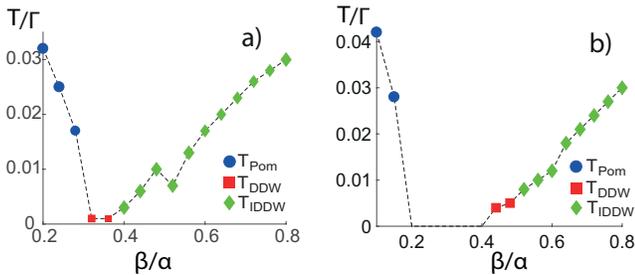}
\caption{Leading particle-hole instability temperatures for SF model. a) $%
a=0.05$, $v_s^2/\protect\alpha=0.05$, $\protect\mu=0.02$,b) $a=0.05$, $v_s^2/%
\protect\alpha=0.05$, $\protect\mu=0.05$. Grey dashed line is guide to the
eye.}
\label{fig3}
\end{figure}

In Fig.\ref{fig3} results of numerical solution of the equations (\ref{app:sf:orders}) are presented for two sets of parameters. The value of parameters characterizing the incommensurability for CDW and IDDW
have been chosen to maximize the critical temperatures. Considering the qualitative character of our approximations for finite $\beta$ below we analyze only the general features of the obtained results.
One can see that the Pomeranchuk instability is considerably more robust to
increasing $\beta $ than is expected from the simplified model (\ref%
{tm:pommucr}). Additionally, the IDDW seems to play a more important
role. Actually, both the results can be qualitatively understood as
being a consequence of the renormalization of the fermionic
self-energy resulting in the replacement $\varepsilon \rightarrow
f(\varepsilon )$. At low frequencies one obtains $\mathrm{Re}%
f(\varepsilon )\sim \varepsilon /Z$, where $Z<1$.Therefore,
the parameter $\mu $ is renormalized to a smaller value $Z\mu $. As
smaller values of $\mu $ qualitatively favor Pomeracnhuk and IDDW
phases, this explains the observed tendency.

As for charge/current order competition one can draw a conclusion that the
boundary between charge and current phases $(\beta /\alpha )_{cr}$ appears
to be remarkably close to the one obtained in the simplified model. The dip
in critical temperatures at intermediate $\beta /\alpha $ is actually also
qualitatively present in the simplified model, however there we concentrated
on competition between phases at a given $T$. One could expect that in this
region superconductivity will re-emerge as a leading instability even if
the remnant Coulomb interaction acts against it. One should also keep in mind that closeness of the different phases
in energy may induce strong fluctuations that can modify the results obtained here in the mean field approximation. These effects, however, should be important only close to phase boundaries. Thus, we suggest that the fluctuations will not change the results qualitatively, but leave detailed investigations for future studies.

\section{Discussion.}

\label{sec:disc}

Considering the obtained non-superconducting phases two tentative scenarios
can be anticipated for the pseudogap state. In the first one, Pomeranchuk
instability is the leading one and is expected to occur at $T_{Pom}\gtrsim
T^{\ast }$. Then, at $T^{\ast }$ the pseudogap would open
due to the formation of an axial CDW\cite{volkov.2016}. However,
the time reversal symmetry breaking does not appear naturally in this
scenario unless more complicated form-factors for the CDW are
considered\cite{wang.2014,gradhand.2015}. Moreover, while in some compounds
\cite{comin.2014} $T_{CDW}$ has been observed to be close to $T^{\ast }$
this does not seem to be the general case\cite{huecker.2014,tabis.2017}.
Additionally, more recent transport data suggests a Fermi surface
reconstruction taking place at $T_{CDW}<T<T^{\ast }$\cite{badoux.2016} to be
distinct from the one caused by the CDW\cite{harrison.2016}.

In the other scenario the leading instability is the DDW that has
its onset at $T^{\ast }$. This is consistent with transport\cite%
{badoux.2016} and ARPES\cite{hashimoto.2010} signatures of the pseudogap. It
has also been shown that an axial d-form factor CDW can emerge on
the Fermi surface reconstructed by the DDW\cite{atkinson.2016,makhfudz.2016}. DDW breaks time reversal symmetry but, as it breaks also
the translational symmetry, additional Bragg peaks at $(\pi
+Q_{x},\pi +Q_{y})$ are expected to appear in the DDW phase. While definitive
experimental evidence for these peaks seems to be lacking\cite{mook.2002,mook.2004,stock.2002,sonier.2009}, we note that the the magnitude of the signal predicted by the BCS-like theories\cite{chakravarty.2001} should change in the case of a strongly frequency-dependent order parameter, such as the one in the SF model. Thus it is possible that the magnitude of the additional peaks predicted using a frequency-independent order parameter could be overestimated. Moreover, the $Q=0$ signal, observed experimentally \cite{fauque.2006,sidis.2013} might orginate from
higher-order processes but this possibility has not been
investigated theoretically sofar.

Additionally, at low dopings there is evidence for a $(\pi+Q_x,\pi+Q_y)$ order not accompanied by the CDW from neutron scattering experiments\cite{haug.2010}. We suggest that this observation can be explained
in terms of the incommensurate IDDW order studied in the present
work. Unlike previous works, where a similar state has been suggested\cite{raczkowski.2007}, in our case IDDW is not accompanied by a charge modulation.

Let us discuss now the values of parameters of
the model (\ref{Mod:lagr}) that might be most suitable for
the cuprates. Relevant values of $\mu$ have been identified in our
previous work\cite{volkov.2016} and are usually of the order of $T^{\ast }$
for the underdoped case. The Fermi surface curvature $\beta /\alpha $
appears now to be another crucial parameter controlling the phase diagram.
One can relate $\beta /\alpha $ to the tight-binding parametrization of the
dispersion in the full Brillouin zone $\varepsilon (\mathbf{p})=-2t(\cos
p_{x}+\cos p_{y})-4t^{\prime }\cos p_{x}\cos p_{y}-2t^{\prime \prime }(\cos
2p_{x}+\cos 2p_{y})$.
\begin{equation*}
\frac{\beta }{\alpha }=\frac{t+2t^{\prime }-4t^{\prime \prime }}{%
t-2t^{\prime }+4t^{\prime \prime }}
\end{equation*}%
Taking result from literature one obtains $\beta /\alpha \approx 0.11$ for
Bi-2201\cite{hashimoto.2008} and $\beta /\alpha \approx 0.15$ for
BSCCO\cite{hoogenboom.2008}. Tight binding fits for YBCO (Ref.%
\onlinecite{nagy.2016} and references therein) yield negative values for $%
\beta /\alpha $ not considered here. At the same time, the
electronic structure around the antinodes in YBCO has also
interpreted\cite{abrikosov.1993} in terms of an extended Van Hove
singularity corresponding to $\beta =0$ case. Moreover, there
are theoretical arguments that such a behavior should be stabilized by
interactions\cite{irkhin.2002}. Anyway, the curvatures are rather
low and seemingly constrain us to the regime where the Pomeranchuk
instability is the leading one. However, drawing quantitative
conclusions about the appearance of the current orders demands
taking into account renormalization of the Fermi surface curvature by the
low-energy interactions, which is beyond the scope of the present
paper.

\section{Conclusion}

\label{sec:concl} We have studied particle-hole instabilities in the
spin-fermion model in the regime where shallowness of the antinodal
dispersion combined with finite AF correlation length leads to a strong
overlap of the 'hot spots' on the Fermi surface. A rich phase diagram has
been obtained as a function of the chemical potential (doping)
relative to the dispersion saddle-points and Fermi surface curvature in the
antinodal regions. The phases obtained include Pomeranchuk and
current-ordered phases previously not encountered in the SF model. We have
shown that for small curvatures $\beta /\xi ^{2}\ll \mu ,T$ an
Eliashberg-like approximation is justified by a small parameter $\sqrt{
[T,\mu ,v_{s}^{2}/\alpha ]/(\alpha /\xi ^{2})}$. The self-energy effects
have been found to promote Pomeranchuk and incommensurate current orders.
The current orders possess attractive features for an explanation of the
pseudogap phase, namely, the particle-hole asymmetric gap in the antinodal
regions, time reversal symmetry breaking and Fermi surface reconstruction
into hole pockets. Moreover, the incommensurate current order obtained in this work can potentially explain the incommensurate magnetism observed at low dopings. Finally, we expect our results to be also of
relevance to other itinerant systems with strong antiferromagnetic
fluctuations.

\begin{acknowledgments}
The authors gratefully acknowledge the financial support of the Ministry of
Education and Science of the Russian Federation in the framework of Increase
Competitiveness Program of NUST~\textquotedblleft MISiS\textquotedblright\
(Nr.~K2-2017-085).
\end{acknowledgments}

\appendix
\begin{widetext}
\section{Matsubara Susceptibilities for Simplified Model}
\label{app:tm}
\subsection{Charge Orders}
In the limit $\beta/\xi^2\to0$ the integral over one of the momenta in (\ref{tm:chi}) yields a factor of $2/\xi$, while for the second one limits can be taken to $\pm \infty$ for $\alpha/\xi^2\gg\mu,\omega_n$. As the resulting sum over $\omega_n$ converges we can neglect the contribution from $\omega_n\gtrsim\alpha/\xi^2$ resulting in (\ref{tm:chicdw0}).

In the opposite case $\beta/\xi^2\gg \omega_n, \mu$ one can extend the integration limits for both momenta to $\pm \infty$. The resulting integrals can be evaluated in this case using:
\[
\frac{1}{a_1 a_2} = \int_{-1/2}^{1/2} dx \frac{1}{[(a_1-a_2)x+(a_1+a_2)/2]^2}.
\]
One has:
\begin{gather*}
(a_1-a_2)x+(a_1+a_2)/2\equiv
(2\alpha p_1 Q_1 - 2\beta p_2 Q_2)x +\alpha (p_1^2+Q_1^2/4) - [\beta(p_2^2+Q_2^2/4)+i \omega+\mu]=\\
\alpha(p_1+Q_1x)^2 - \beta(p_2+Q_2x)^2-(i\omega+\mu+(\alpha Q_1^2-\beta Q_2^2)(x^2-1/4)).
\end{gather*}
The resulting integrals over momenta converge for all $x$. Consequently one can exchange the integration order to obtain:
\begin{gather*}
- \int_{-1/2}^{1/2} dx \int_{-\infty}^{\infty} d p_2 \int_{-\infty}^{\infty} d p_1
\frac{1}{\{\alpha p_1^2 - \beta p_2^2-[i\omega+\mu+\gamma(4x^2-1)]\}^2}=
\\
\int_{-1/2}^{1/2} dx \int_{-\infty}^{\infty} d p_2
\frac{i\pi{\rm sgn} [\omega_n]}{2\sqrt{\alpha}}
\frac{1}{[\beta p_2^2+i\omega+\mu+\gamma(4x^2-1)]^{3/2}} =
\\
\int_{-1/2}^{1/2} dx
\frac{i\pi{\rm sgn} [\omega_n]}
{
\sqrt{\alpha\beta}
[i\omega+\mu+\gamma(4x^2-1)]
}
=
\frac{- i\pi{\rm sgn} [\omega_n]
{\rm arctanh}
\sqrt{
\frac{\gamma}
{\gamma-i\omega-\mu}
}
}
{
\sqrt{\alpha\beta}
\sqrt{\gamma}
\sqrt{\gamma-i\omega-\mu}
}.
\end{gather*}
The sum over Matsubara frequencies appears to diverge at large $\omega_n$, but only logarithmically. This allows one to obtain the leading contribution to $\chi_{CDW}({\bf Q})$ by introducing a cutoff at $\omega_n\sim\beta/\xi^2$ in the sum and neglecting the region $\omega_n\gtrsim\beta/\xi^2$ provided that $\beta/\xi^2\gg\mu,T$.

To study the stability of the ${\bf Q}=0$ phase at finite $\beta$ we expand the CDW susceptibility in Eq.(\ref{tm:chi}) in powers of ${\bf Q}$. One obtains that $\left.\frac{\partial \chi_i}{\partial Q_j}\right|_{{\bf Q} = 0}$ as well as$\left.\frac{\partial^2 \chi_i}{\partial Q_1 \partial Q_2}\right|_{{\bf Q} = 0}$ vanish. Since $\bf Q$ is along the diagonal the condition for the critical value of $\mu$ is $\left.\frac{\partial^2 \chi_1}{\partial Q_1^2}+\frac{\partial^2 \chi_2}{\partial Q_1^2}\right|_{{\bf Q} = 0}=\left.\frac{\partial^2 \chi_1}{\partial Q_2^2}+\frac{\partial^2 \chi_2}{\partial Q_2^2}\right|_{{\bf Q} = 0}=0$. Performing the expansion and integrating over momenta ($\alpha\xi^2\gg\mu,\omega_n$ is assumed)one obtains:
\begin{equation}
\begin{gathered}
\frac{\alpha Q^2}{4}
\nu_0 T \sum_{\omega_n} \frac{i \pi \; {\rm sgn}[\omega_n]}{3\sqrt{\alpha\beta}}
\frac{3(i \omega_n+\mu)+2\beta/\xi^2}
{
(i \omega_n+\mu)^2
}
\sqrt{\frac{\beta /\xi^2}{(\beta/\xi^2+i \omega_n+\mu)^3}}
\\
-\frac{\beta Q^2}{4}
\nu_0 T \sum_{\omega_n} \frac{i \pi \; {\rm sgn}[\omega_n]}{6\sqrt{\alpha\beta}}
\frac
{
\sqrt{\beta/\xi^2} (9 (i \omega_n+\mu)^2 +10(i \omega_n+\mu) \beta/\xi^2 + 4 (\beta/\xi^2)^2)
}
{
(i \omega_n+\mu )^2 (i \omega_n+\mu + \beta/\xi^2)^{(5/2)}
}=0.
\end{gathered}
\label{app:pommucr}
\end{equation}
Expanding this result for $\mu\ll T$ one obtains (\ref{tm:pommucr}) after summation.

\subsection{Current Orders}
For ${\bf Q}=0$ we rewrite the expression in (\ref{tm:chi})
\begin{gather*}
\frac{1}
{
(i \omega_n - \alpha p_1^2 +\beta p_2^2 +\mu_1)
(i \omega_n - \alpha p_2^2 +\beta p_1^2 -\mu_2)
}
=
\\
\frac{1}{\alpha+\beta}
\frac{1}
{(\alpha-\beta) p_2^2-(i\omega_n+\mu_-)}
\left\{
\frac{\alpha}{\alpha p_1^2 - (\beta p_2^2 +i \omega_n + \mu_1)}
-
\frac{\alpha-\beta}
{(\alpha-\beta) p_1^2 -(i \omega+ \mu_+)}
\right\}
\\
+
\frac{\alpha}{\alpha+\beta}
\frac{1}
{
    [(\alpha-\beta) p_1^2 -(i \omega+ \mu_+)]
    [\alpha p_2^2-(i \omega_n+\beta p_1^2-\mu_2)]
}.
\end{gather*}
Assuming only $\alpha/\xi^2\gg\mu,\omega_n$ one can evaluate the integral over momenta analytically. This yields
\begin{equation}
\begin{gathered}
\chi_D(T) = T \sum_{\omega_n}
\frac{
    4 \pi i \;{\rm sgn} [\omega_n]
    {\rm arctanh}
    \left\{
        \sqrt{\frac{\alpha/\xi^2}{\beta /\xi^2+i \omega_n + \mu}}
    \right\}
}
{(\alpha+\beta)(i \omega_n + \mu)}
+
\frac{4}{\alpha+\beta}
\frac{
{\rm arctanh}
\left\{
\sqrt{\frac{(\alpha-\beta)/\xi^2}{i \omega_n+\mu}}
\right\}
{\rm arctanh}
\left\{
\sqrt{\frac{(\alpha-\beta)/\xi^2}{i \omega_n+\mu}}
\right\}
}
{i \omega_n + \mu}.
\end{gathered}
\end{equation}
From this equation one can obtain (\ref{tm:chiddw0}) and (\ref{tm:chiddw}). For finite ${\bf Q}$ a closed form for $\chi_D({\bf Q})$ can be obtained for $\beta/\xi^2\gg T,\mu$ using:
\[
\frac{1}{a_1 a_2} = \int_{-1/2}^{1/2} dx \frac{1}{[(a_1-a_2)x+(a_1+a_2)/2]^2}.
\]
For the momentum integral in (\ref{tm:chi}) we obtain:
\begin{gather*}
-\int_{-\infty}^{\infty} d p_1 \int_{-\infty}^{\infty}d p_2 \int_{-1/2}^{1/2} dx \frac{1}{[...]^2},
\\
[...]=
\left[(\alpha+\beta)x+\frac{\alpha-\beta}{2}\right](p_1^2+Q_1^2/4)
+
\left[(\alpha-\beta)x+\frac{\alpha+\beta}{2}\right] p_1 Q_1
\\
+
\left[-(\alpha+\beta)x+\frac{\alpha-\beta}{2}\right](p_2^2+Q_2^2/4)
+
\left[(\alpha-\beta)x-\frac{\alpha+\beta}{2}\right] p_2 Q_2
-(i \omega + \mu).
\end{gather*}
To change the integration order we need to assume $Q_1,Q_2\neq0$ as $\pm(\alpha+\beta)x+(\alpha-\beta)/2$ can vanish inside the $x$ integration region while $(\alpha-\beta)x\pm(\alpha+\beta)/2$ does not cross zero for all $x$. To integrate over $p_1$ we rewrite $[...]$:
\begin{gather*}
[...]=
\left[(\alpha+\beta)x+\frac{\alpha-\beta}{2}\right]
\left(
p_1+\frac{(\alpha-\beta)x+(\alpha+\beta)/2}{(\alpha+\beta)x+(\alpha-\beta)/2}\frac{Q_1}{2}
\right)^2
+
\frac{Q_1^2}{4}\frac{4 \alpha \beta (x^2-1/4)}{(\alpha+\beta)x+(\alpha-\beta)/2}
\\
-
\left[(\alpha+\beta)x-\frac{\alpha-\beta}{2}\right]
\left(
p_2+\frac{-(\alpha-\beta)x+(\alpha+\beta)/2}{(\alpha+\beta)x-(\alpha-\beta)/2}\frac{Q_2}{2}
\right)^2
-\frac{Q_2^2}{4}\frac{4 \alpha \beta (x^2-1/4)}{(\alpha+\beta)x-(\alpha-\beta)/2}
-(i \omega + \mu).
\end{gather*}
Now one can simplify the calculation by shifting the integration variables. First let us integrate over $p_1$. The answer depends on the sign of $(\alpha+\beta)x+\frac{\alpha-\beta}{2}$:
\begin{gather*}
\frac{-1}{2\left(
\left[(\alpha+\beta)x-\frac{\alpha-\beta}{2}\right]p_2^2+
\frac{Q_2^2 \alpha \beta (x^2-1/4)}{(\alpha+\beta)x-(\alpha-\beta)/2}
-\frac{Q_1^2 \alpha \beta (x^2-1/4)}{(\alpha+\beta)x+(\alpha-\beta)/2}
+i\omega+\mu
\right)^{(3/2)}}
\cdot
\\
\cdot
\begin{cases}
\frac{-i \pi\; {\rm sgn} [\omega_n]}{\sqrt{(\alpha+\beta)x+\frac{\alpha-\beta}{2}}}
 &\mbox{if } x>-\frac{\alpha-\beta}{2(\alpha+\beta)} \\
\frac{\pi}{\sqrt{-(\alpha+\beta)x-\frac{\alpha-\beta}{2}}}
 & \mbox{if }x<-\frac{\alpha-\beta}{2(\alpha+\beta)} \end{cases}.
\end{gather*}

The remaining integral over $p_2$ yields:
\begin{gather*}
\int_{-\infty}^{\infty} d p_2
\frac{1}{\left(
\left[(\alpha+\beta)x-\frac{\alpha-\beta}{2}\right]p_2^2+
\frac{Q_2^2 \alpha \beta (x^2-1/4)}{(\alpha+\beta)x-(\alpha-\beta)/2}
-\frac{Q_1^2 \alpha \beta (x^2-1/4)}{(\alpha+\beta)x+(\alpha-\beta)/2}
+i\omega+\mu
\right)^{(3/2)}}
=
\\
=
\frac{1}{
i\omega+\mu+
\frac{Q_2^2 \alpha \beta (x^2-1/4)}{(\alpha+\beta)x-(\alpha-\beta)/2}
-\frac{Q_1^2 \alpha \beta (x^2-1/4)}{(\alpha+\beta)x+(\alpha-\beta)/2}
}
\cdot
\begin{cases}
\frac{2}{\sqrt{(\alpha+\beta)x-\frac{\alpha-\beta}{2}}}
 &\mbox{if }x>\frac{\alpha-\beta}{2(\alpha+\beta)} \\
\frac{-2 i {\rm sgn} [\omega_n]}{\sqrt{-(\alpha+\beta)x+\frac{\alpha-\beta}{2}}}
 & \mbox{if }x<\frac{\alpha-\beta}{2(\alpha+\beta)} \end{cases}.
\end{gather*}
Combining the results above one obtains two contributions. The first one is:
\begin{gather*}
I_1=\int_{-\frac{\alpha-\beta}{2(\alpha+\beta)}}^{\frac{\alpha-\beta}{2(\alpha+\beta)}} dx
\frac{\pi}{\sqrt{(\alpha-\beta)^2/4-(\alpha+\beta)^2x^2}}
\\
\frac{(\alpha+\beta)^2x^2-(\alpha-\beta)^2/4}
{(i\omega+\mu)[(\alpha+\beta)^2x^2-(\alpha-\beta)^2/4]
+\alpha\beta(x^2-1/4)\{(Q_2^2-Q_1^2)(\alpha+\beta)x
+(Q_2^2+Q_1^2)(\alpha-\beta)/2\}}.
\end{gather*}
Or, after a change of variables $x\to(\alpha-\beta)x/(\alpha+\beta)/2$ and some algebra:
\[
\int_{-1}^{1} dx
\frac{\pi}{\alpha+\beta}
\frac{\sqrt{1-x^2}}
{(i\omega+\mu)[1-x^2]
+\frac{\alpha\beta}{\alpha-\beta}
\left(1-x^2\left(\frac{\alpha-\beta}{\alpha+\beta}\right)^2\right)\{\delta Q^2x
+Q^2\}},
\]
where $Q^2=(Q_1^2+Q_2^2)/2$ and $\delta Q^2 = (Q_2^2-Q_1^2)/2$. For $\delta Q^2$ = 0 one can evaluate the integral analytically to obtain
\[
I_1(Q,Q)=\frac{\pi^2}{\alpha+\beta}
\frac{1}{i\omega+\mu+\frac{\alpha\beta (\alpha-\beta)Q^2}{(\alpha+\beta)^2}}
    \left[
    1-
    \sqrt{
    \frac{4\alpha^2\beta^2Q^2/(\alpha+\beta)^2/(\alpha-\beta)}
    {i\omega+\mu
    +\frac{\alpha\beta Q^2}{\alpha-\beta}
    }
    }
    \right].
\]
The second contribution is:
\begin{gather*}
I_2=\int_{\frac{\alpha-\beta}{2(\alpha+\beta)}}^{\frac{1}{2}}
+\int_{-\frac{1}{2}}^{-\frac{\alpha-\beta}{2(\alpha+\beta)}}
dx
\frac{i \pi {\rm sgn} [\omega_n]}{\sqrt{(\alpha+\beta)^2x^2-(\alpha-\beta)^2/4}}
\\
\frac{(\alpha+\beta)^2x^2-(\alpha-\beta)^2/4}
{(i\omega+\mu)[(\alpha+\beta)^2x^2-(\alpha-\beta)^2/4]
+\alpha\beta(x^2-1/4)\{2\delta Q^2(\alpha+\beta)x
+Q^2(\alpha-\beta)\}}=
\\
\int_{1}^{\frac{\alpha+\beta}{\alpha-\beta}}
+\int_{-\frac{\alpha+\beta}{\alpha-\beta}}^{-1}
\frac{dx}{\alpha+\beta}
\frac{i \pi {\rm sgn} [\omega_n]}{\sqrt{x^2-1}}
\frac{1}
{i\omega+\mu
+
\frac{
    \alpha\beta[(\alpha-\beta)^2x^2-(\alpha+\beta)^2]
    \{\delta Q^2x +Q^2\}
}
{
    (\alpha-\beta)(\alpha+\beta)^2(x^2-1)
}
}.
\end{gather*}
For $\delta Q^2=0$ the integral in $I_2$ can be evaluated to obtain
\[
\frac{4 i \pi {\rm sgn} [\omega_n] }
{(\alpha+\beta) \left[ i\omega+\mu +\frac{\alpha(\alpha-\beta)\beta Q^2}{(\alpha+\beta)^2}\right]}
\left(
{\rm arctanh} \sqrt{\frac{\beta}{\alpha}}
-
\sqrt{
\frac{\frac{\alpha\beta}{(\alpha+\beta)^2}\frac{\alpha\beta Q^2}{\alpha-\beta}}{i\omega+\mu +\frac{\alpha\beta Q^2}{\alpha-\beta}}
}
\left(
\frac{i \pi {\rm sgn} [\omega_n]}{2}+
{\rm arctanh}
\sqrt{
\frac{\frac{\alpha\beta Q^2}{\alpha-\beta}}{i\omega+\mu +\frac{\alpha\beta Q^2}{\alpha-\beta}}
}
\right)
\right)
\]
Combining this with the first contribution we get:
\begin{equation}
\begin{gathered}
\chi_{DDW}(Q,Q)=
\frac{\pi^2}{\alpha+\beta}
\frac{1}{i\omega+\mu+\frac{\alpha\beta (\alpha-\beta)Q^2}{(\alpha+\beta)^2}}
+
\\\frac{4 i \pi {\rm sgn} [\omega_n] }
{(\alpha+\beta) \left[ i\omega+\mu +\frac{\alpha(\alpha-\beta)\beta Q^2}{(\alpha+\beta)^2}\right]}
\left(
{\rm arctanh} \sqrt{\frac{\beta}{\alpha}}
-
\sqrt{
\frac{\frac{\alpha\beta}{(\alpha+\beta)^2}\frac{\alpha\beta Q^2}{\alpha-\beta}}{i\omega+\mu +\frac{\alpha\beta Q^2}{\alpha-\beta}}
}
{\rm arctanh}
\sqrt{
\frac{\frac{\alpha\beta Q^2}{\alpha-\beta}}{i\omega+\mu +\frac{\alpha\beta Q^2}{\alpha-\beta}}
}
\right).
\label{app:chiddiag}
\end{gathered}
\end{equation}
The expression (\ref{app:chiddiag}) can be used to calculate the IDDW/DDW phase boundary because for small ${\bf Q}$ one can show that $\chi_D({\bf Q}\to 0)\approx a+b(Q_x^2+Q_y^2)$. From the condition $\frac{\partial^2 \chi_D}{\partial Q_1^2}=0$ we get after evaluation of the Matsubara sum:
\begin{equation}
\begin{gathered}
-\frac{i}{2 \pi (\alpha+\beta)^2}
\left[
\frac{(\alpha\beta)^{3/2}}{\alpha-\beta}+\frac{\alpha \beta (\alpha-\beta)}{\alpha+\beta}
{\rm arctanh}\left[\sqrt{\frac{\beta}{\alpha}}\right]
\right]
\left(
\psi'\left[\frac{1}{2}+\frac{i \mu}{2 \pi T}\right]-\psi'\left[\frac{1}{2}-\frac{i \mu}{2 \pi T}\right]\right)
\\
+
\frac{(\alpha - \beta)\alpha \beta \pi^2}{8 (\alpha + \beta)^3 \cosh^2[\mu/2T]}=0.
\end{gathered}
\label{app:iddwcr}
\end{equation}
This expression has been used to calculate the IDDW/DDW boundary numerically. As is evident from Fig. \ref{Fig:tmphasediagr} $\mu/T$ becomes small for $(\alpha-\beta)\ll\alpha$. Expanding (\ref{app:iddwcr}) for $\mu/T\ll1,(\alpha-\beta)\ll\alpha$ one obtains Eq. \ref{tm:ddwmucr}.

For numerical calculations of $\chi_{DDW}$ and orientation of ${\bf Q}$ we have evaluated the Matsubara sum before the integrals in $I_1,I_2$. Using $T\sum\frac{1}{i\omega+a}=\tanh(a/2T)/2$ one obtains for $I_1$:
\begin{equation}
\chi_{DDW}^1=\frac{\pi}{2(\alpha+\beta)}
\int_{-1}^{1} \frac{dx}{\sqrt{1-x^2}}
\tanh\left\{
\frac{\mu}{2T}+
\frac{\alpha}{\alpha-\beta}
\frac{1-\left(\frac{\alpha-\beta}{\alpha+\beta}\right)^2x^2}{1-x^2}
\frac{\beta Q^2+\beta \delta Q^2 x}{2 T}
\right\}.
\label{app:chidnum1}
\end{equation}
Calculating the derivative $\frac{d \chi_{DDW}^1}{ d \delta Q^2}$ one obtains that it is negative for $\delta Q^2>0$ and positive for $\delta Q^2<0$. Consequently, $J_1$ has a global maximum at $\delta Q^2=0$. Moreover, one can see that $\chi_{DDW}^1$ increases with $Q^2$. For $I_2$ the resulting Matsubara sum diverges logarithmically, however one can subtract the divergent part $T\sum\frac{1}{|\omega|}$. Using
$T\sum\frac{i {\rm sgn} [\omega_n]}{i\omega+a}- \frac{1}{|\omega|} =
\left[2 \psi(0, \frac{1}{2}) - \psi\left(0, \frac{1}{2}+\frac{ia}{2\pi T}\right) -
  \psi\left(0, \frac{1}{2}-\frac{ia}{2\pi T}\right)
  \right]/2\pi$ one obtains:
\begin{equation}
\begin{gathered}
\chi_{DDW}^2=
\frac{1}{2(\alpha+\beta)}
\int_{1}^{\frac{\alpha+\beta}{\alpha-\beta}}
+\int_{-\frac{\alpha+\beta}{\alpha-\beta}}^{-1}
\frac{dx}{\sqrt{x^2-1}}
\left[
2 \psi\left(0, \frac{1}{2}\right)
- \psi\left(0, \frac{1}{2}+\frac{i\kappa}{2\pi T}\right)
- \psi\left(0, \frac{1}{2}-\frac{i\kappa}{2\pi T}\right)
\right]
,
\\
\kappa=
\mu
+
\frac{
    \alpha\beta[(\alpha-\beta)^2x^2-(\alpha+\beta)^2]
    \{\delta Q^2x +Q^2\}
}
{
    (\alpha-\beta)(\alpha+\beta)^2(x^2-1)
}.
\end{gathered}
\label{app:chidnum2}
\end{equation}
The divergent part is then evaluated with a cutoff at $\omega=\beta/\xi^2$. The expressions (\ref{app:chidnum1}),(\ref{app:chidnum2}) have been used to calculate $\chi_{DDW}({\bf Q})$ numerically:
\begin{equation}
\chi_{DDW}=\chi_{DDW}^1+\chi_{DDW}^2+
\frac{4 {\rm arctanh} \sqrt{\frac{\beta}{\alpha}}}{\alpha+\beta}
\left(\psi\left(0, 1+\frac{\beta/\xi^2}{2\pi T}\right)-\psi(0, 1/2)\right)
\label{app:chidnum}
\end{equation}

Orientation and magnitude of ${\bf Q}$ are found maximizing $\chi_{DDW}({\bf Q})$. One can get some analytical insight on the possible orientation of ${\bf Q}$ in the case $\beta\ll\alpha$. $I_2$ can be approximately evaluated taking $x\approx\pm \frac{\alpha-\beta}{2(\alpha+\beta)}$ in the integral:
\begin{gather*}
I_2\approx
\frac{2 i \pi {\rm sgn} [\omega_n] \sqrt{\frac{\beta}{\alpha}}}
{\alpha \left[ i\omega+\mu +\beta Q_1^2\right]}
\left(
1
-
\sqrt{
\frac{\beta Q_1^2}
{i\omega+\mu +\beta Q_1^2}
}
{\rm arctanh}
\sqrt{
\frac{i\omega+\mu +\beta Q_1^2}
{\beta Q_1^2}
}
\right)
\\
+
\frac{2 i \pi {\rm sgn} [\omega_n] \sqrt{\frac{\beta}{\alpha}}}
{\alpha \left[ i\omega+\mu +\beta Q_2^2\right]}
\left(
1
-
\sqrt{
\frac{\beta Q_2^2}
{i\omega+\mu +\beta Q_2^2}
}
{\rm arctanh}
\sqrt{
\frac{i\omega+\mu +\beta Q_2^2}
{\beta Q_2^2}
}
\right)
\end{gather*}
At $T,\mu\ll\beta Q^2$ one can go to integration over $\omega$. The second terms in the brackets yield after integration over $\omega$: $\sqrt{\beta/\alpha^3}{\rm const} + O(\mu/\beta Q^2)$ , consequently:
\begin{gather*}
\chi^2_{DDW}\approx
\int \frac{d\omega}{2\pi}
\frac{2 i \pi {\rm sgn} [\omega_n] \sqrt{\frac{\beta}{\alpha}}}
{\alpha \left[ i\omega+\mu +\beta Q_2^2\right]}+
\frac{2 i \pi {\rm sgn} [\omega_n] \sqrt{\frac{\beta}{\alpha}}}
{\alpha \left[ i\omega+\mu +\beta Q_2^2\right]}
+\sqrt{\beta/\alpha^3}{\rm const} + O(\mu/\beta Q^2)
\sim
\\
2\sqrt{\beta/\alpha^3}\log\{(\beta/\xi^2)^2/[(\beta Q^2 +\mu)^2-(\beta \delta Q^2)^2]\} + O(\mu/\beta Q^2)+{\rm const}.
\end{gather*}
$\chi^1_{DDW}({\bf Q})$ can be shown to be bounded from above by $\frac{\pi^2}{2(\alpha+\beta)}$. Consequently, for $\beta /\xi^2\gg\beta Q^2\gg\mu,T$ contribution from $\chi_{DDW}^2$ is dominant and maximizing $\delta Q^2$ is favorable. The maximal absolute value of $\delta Q^2$ is $Q^2$ that is reached is reached for ${\bf Q}$ along one of the BZ axes. However as $\beta$ decreases $\chi_{DDW}^2$ eventually becomes less important due to the factor $\sqrt{\beta/\alpha}$. As $\chi_{DDW}^1$ has been shown above to be maximal for $\delta Q^2=0$ one expects a transition to diagonal ${\bf Q}$ at low $\beta/\alpha$. This is in line with the results of numerical calculation in Fig. \ref{Fig:tmqddw}.

\section{Expression for the current density}
\label{app:cur}
The definition of the current density directly follows from the standard expression for the magnetic part of the action $S_{\mathrm{mag}}$ in the presence of a vector potential $\mathbf{A}\left( \mathbf{r}\right) $. Here we derive the bare part $S_{\mathrm{0}}\left[ \mathbf{A}\right] $ of the action $S$ in the presence of an external potential $\mathbf{A}\left( \mathbf{r}\right) $ and derive the expression for the current using a standard formula from electrodynamics for the magnetic part $S_{\mathrm{mag}}$ of the action
\begin{equation}
S_{\mathrm{mag}}=-\frac{1}{c}\int \mathbf{j}\left( \tau ,\mathbf{r}\right)
\mathbf{A}\left( \tau ,\mathbf{r}\right) d\tau d\mathbf{r,}  \label{r12}
\end{equation}%
For simplicity we consider the energy operator to be of the form%
\begin{equation}
\hat{\varepsilon}\left( \mathbf{-}i\mathbf{\nabla }\right) =J\left( 2-\cos
\left( \mathbf{-}i\mathbf{\nabla a}_{x}\right) -\cos \left( \mathbf{-}i%
\mathbf{\nabla a}_{y}\right) \right) .  \label{r1}
\end{equation}%
In Eq. (\ref{r1}) $\mathbf{a}_{x}$ and $\mathbf{a}_{y}$ are lattice vectors
directed along $x$ and $y$ bonds of the $CuO$ lattice, respectively, and $%
\left\vert \mathbf{a}_{x}\right\vert =\left\vert \mathbf{a}_{y}\right\vert
=a_{0}.$

The energy operator for the system with the vector potential can easily be
written using the minimal coupling equivalent to the Peierls' substitution in Eq. (%
\ref{r1})
\begin{equation}
-i\mathbf{\nabla \rightarrow }-i\mathbf{\nabla -}\frac{e}{c}\mathbf{A}
\label{d1}
\end{equation}%
Then, the energy operator takes the form%
\begin{eqnarray}
&&\hat{\varepsilon}\left( \mathbf{-}i\mathbf{\nabla -}\frac{e}{c}\mathbf{A}%
\left( \mathbf{r}\right) \right) =-J\Big[\cos \left( \left( \mathbf{-}i%
\mathbf{\nabla -}\frac{e}{c}\mathbf{\mathbf{A}\left( \mathbf{r}\right) }%
\right) \mathbf{a}_{x}\right)  \notag \\
&&\cos \left( \left( \mathbf{-}i\mathbf{\nabla -}\frac{e}{c}\mathbf{\mathbf{A%
}\left( \mathbf{r}\right) }\right) \mathbf{a}_{y}\right) -2\Big]  \label{d2}
\end{eqnarray}%
The current operator should be defined calculating the linear term of the
expansion of the action $S_{\mathrm{mag}}$ in $\mathbf{\mathbf{A}\left(
\mathbf{r}\right) }$.

As the vector potential $\mathbf{A}\left( \mathbf{r}\right) $ does not
commute with the gradient, the expansion in $\mathbf{A}\left( \mathbf{r}%
\right) $ is not trivial and we use time ordering products. For any
non-commuting operators $A$ and $B$ one has
\begin{eqnarray}
&&\exp \left( A+B\right) =\exp AT_{\alpha }\exp \left( \int_{0}^{1}\tilde{B}%
\left( \alpha \right) d\alpha \right)  \notag \\
&=&T_{\alpha }^{A}\exp \left( \int_{0}^{1}\tilde{B}^{A}\left( \alpha \right)
d\alpha \right) \exp A,  \label{d3}
\end{eqnarray}%
where
\begin{equation}
\tilde{B}\left( \alpha \right) =e^{-A\alpha }Be^{A\alpha },\quad \tilde{B}%
^{A}\left( \alpha \right) =e^{A\alpha }Be^{-A\alpha },  \label{d4}
\end{equation}%
and $T_{\alpha }$ and $T_{\alpha }^{A}$ are time ordering and anti time
ordering operators, respectively.

Introducing operators
\begin{eqnarray}
\mathbf{\tilde{A}}\left( \mathbf{r,}\alpha \right) &=&e^{-\alpha \mathbf{a}%
_{x}\mathbf{\nabla }}\mathbf{A}\left( \mathbf{r}\right) e^{\alpha \mathbf{a}%
_{x}\mathbf{\nabla }}=\mathbf{A}\left( \mathbf{r-}\alpha \mathbf{a}%
_{x}\right) ,  \notag \\
\quad \mathbf{\tilde{A}}^{A}\left( \mathbf{r,}\alpha \right) &=&e^{\alpha
\mathbf{a}_{x}\mathbf{\nabla }}\mathbf{A}\left( \mathbf{r}\right) e^{-\alpha
\mathbf{a}_{x}\mathbf{\nabla }}=\mathbf{A}\left( \mathbf{r+}\alpha \mathbf{a}%
_{x}\right)  \notag \\
&&  \label{d5}
\end{eqnarray}%
one comes to the following expressions
\begin{eqnarray}
&&\exp \left[ \pm \left( \mathbf{\nabla -}\frac{ie}{c}\mathbf{\mathbf{A}%
\left( \mathbf{r}\right) }\right) \mathbf{a}_{x}\right]  \label{d6} \\
&=&\exp \left[ \pm \mathbf{a}_{x}\mathbf{\nabla }\right] \exp \left[ \mp
\frac{ie}{c}\int_{0}^{1}\mathbf{a}_{x}\mathbf{A}\left( \mathbf{r\mp }\alpha
\mathbf{a}_{x}\right) d\alpha \right]  \notag \\
&=&\exp \left[ \mp \frac{ie}{c}\int_{0}^{1}\mathbf{a}_{x}\mathbf{A}\left(
\mathbf{r\pm }\alpha \mathbf{a}_{x}\right) d\alpha \right] \exp \left[ \pm
\mathbf{a}_{x}\mathbf{\nabla }\right]  \notag
\end{eqnarray}%
Using Eq. (\ref{d6}) we rewrite the energy operator $\hat{\varepsilon}\left(
\mathbf{-}i\mathbf{\nabla -}\frac{e}{c}\mathbf{A}\left( \mathbf{r}\right)
\right) $. This leads to to the
following expression for the action
\begin{eqnarray}
&&S_{\mathrm{0}}\left[ \mathbf{A}\right] =-t\int c^\dagger\left( \tau ,%
\mathbf{r}\right) \Big[e^{-\frac{ie}{c}\int_{0}^{1}\mathbf{a}_{x}\mathbf{A}%
\left( \mathbf{r+}\alpha \mathbf{a}_{x}\right) d\alpha }c \left( \tau ,%
\mathbf{r+a}_{x}\right)  \notag \\
&&+e^{\frac{ie}{c}\int_{0}^{1}\mathbf{a}_{x}\mathbf{A}\left( \mathbf{r-}%
\alpha \mathbf{a}_{x}\right) d\alpha } c \left( \tau ,\mathbf{r-a}%
_{x}\right) -2 c \left( \tau ,\mathbf{r}\right) \Big]d\tau d\mathbf{r}
\label{d6a}
\end{eqnarray}

Now, expanding the exponentials in the vector potential $\mathbf{A}\left(
\mathbf{r}\right) ,$ and comparing the linear in $\mathbf{A}\left( \mathbf{r}%
\right) $ term with $S_{\mathrm{mag}}$, Eq. (\ref{r12}), we bring the
correlation function for the current density to the form%
\begin{eqnarray}
&&\mathbf{j}_{x,y}\left( \tau ,\mathbf{r}\right) =-\frac{i}{2}eJ\mathbf{a}%
_{x,y}  \notag \\
&&\times \int_{0}^{1}\Big[\left\langle c^\dagger\left( \tau ,\mathbf{r-}%
\alpha \mathbf{a}_{x,y}\right) c \left( \tau ,\mathbf{r}+\left( 1-\alpha
\right) \mathbf{a}_{x,y}\right) \right\rangle  \notag \\
&&-\left\langle c^\dagger\left( \tau ,\mathbf{r+}\alpha \mathbf{a}%
_{x,y}\right) c \left( \tau ,\mathbf{r-}\left( 1-\alpha \right) \mathbf{a}%
_{x,y}\right) \right\rangle \Big]d\alpha ,  \label{d7}
\end{eqnarray}%
where $\mathbf{j}_{x,y}$ are $x$- and $y$- components of the current
density. The angular brackets in Eq. (\ref{d7}) stand for averaging with the
action $S$ of the system. We will use for this averaging the action in the
mean field approximation.

We emphasize that the current density $\mathbf{j}_{x,y}\left( \tau ,\mathbf{r%
}\right) $ is a function of the continuous coordinate $\mathbf{r}$ and Eq. (%
\ref{d7}) is valid not only on the sites of the lattice. This is very
important because in some cases non-zero circulating currents turn to zero
at these points.

In order to calculate physical quantities can expand the fields $c \left(
\mathbf{r}\right) $ in the Bloch functions $\psi _{\mathbf{P}}$
\begin{equation}
c \left( \mathbf{r}\right) =\int c _{\mathbf{P}}\psi _{\mathbf{P}%
}\left( \mathbf{r}\right) \frac{d\mathbf{P}}{\left( 2\pi \right) ^{2}},
\label{d8}
\end{equation}%
where $\psi _{\mathbf{P}}\left( \mathbf{r}\right) $ has the standard form
\begin{equation}
\psi _{\mathbf{p}}\left( \mathbf{r}\right) =e^{i\mathbf{Pr}}u_{\mathbf{p}%
}\left( \mathbf{r}\right) ,  \label{d9}
\end{equation}%
$u_{\mathbf{p}}\left( \mathbf{r}\right) $ is periodic function with the
period $\mathbf{a}_{x,y}$ and $\mathbf{P}$ is a quasimomentum in the first
Brillouin zone. Note that in the main text the quasimomenta are defined in units of the inverse lattice spacing.

However, as we use the spectrum, Eq. (\ref{d2}), corresponding to a tight
binding limit, the eigenfunctions of the Hamiltonian are localized near the
lattice sites and it is convenient to expand the Bloch functions in Wannier
functions $w_{\mathbf{R}_{n}}\left( \mathbf{r}\right) $ representing the
functions $\psi _{\mathbf{P}}\left( \mathbf{r}\right) $ as
\begin{equation}
\psi _{\mathbf{P}}\left( \mathbf{r}\right) =\sum_{\mathbf{R}_{n}}e^{i\mathbf{%
PR}_{\mathbf{n}}}w_{\mathbf{R}_{\mathbf{n}}}\left( \mathbf{r}\right) ,
\label{d10}
\end{equation}%
where $\mathbf{R}_{\mathbf{n}}\mathbf{=a}_{x}n_{x}+\mathbf{a}_{y}n_{y},$ $%
\mathbf{n=}\left( n_{x},n_{y}\right) $, $n_{x},n_{y}=0,\pm 1,\pm 2,\pm 3....$
and $N$ is the total number of the sites. Then,%
\begin{equation}
c \left( \mathbf{r}\right) =\sum_{\mathbf{R}_{\mathbf{n}}}\int c_{%
\mathbf{P}}e^{i\mathbf{PR}_{\mathbf{n}}}w_{\mathbf{R}_{\mathbf{n}}}\left(
\mathbf{r}\right)  \label{d11}
\end{equation}

The functions $w_{\mathbf{R}_{n}}\left( \mathbf{r}\right) $ are localized
near the sites with the coordinates $\mathbf{r-R}_{\mathbf{n}}$. The
function $w_{0}\left( \mathbf{r}\right) $ is localized near $\mathbf{r}=0$,
and $w_{\mathbf{R}_{\mathbf{n}}}\left( \mathbf{r}\right) =w_{0}\left(
\mathbf{r-R}_{\mathbf{n}}\right) $. The Wannier functions are normalized as
follows
\begin{equation}
\int \left\vert w_{\mathbf{R}_{\mathbf{n}}}\left( \mathbf{r}\right)
\right\vert ^{2}d\mathbf{r}=1  \label{d12}
\end{equation}

Taking the Fourier transform of the current%
\begin{equation}
\mathbf{j}_{x,y}\left( \tau ,\mathbf{q}\right) =\int \mathbf{j}_{x,y}\left(
\tau ,\mathbf{r}\right) e^{-i\mathbf{qr}}d\mathbf{r}  \label{d13}
\end{equation}%
we substitute Eq. (\ref{d11}) into Eq. (\ref{d7}) and the latter into Eq. (%
\ref{d13}). The we shift $\mathbf{r\rightarrow r-}\alpha \mathbf{a}_{x,y}$
in the first term in $\mathbf{j}_{x,y}\left( \mathbf{r}\right) $ and $%
\mathbf{r\rightarrow r}+\alpha \mathbf{a}_{x,y}$ in the second one and use
the fact that the product $w_{\mathbf{R}_{\mathbf{n}_{1}}}\left( \mathbf{r}%
\right) w_{\mathbf{R}_{n_{2}}}\left( \mathbf{r\pm a}_{x,y}\right) $ is
essentially different from zero only for $\mathbf{R}_{\mathbf{n}_{1}}=%
\mathbf{R}_{\mathbf{n}_{2}}\mp \mathbf{a}_{x,y}$. The the integral over $%
\mathbf{r}$ reduces to the following expression%
\begin{equation}
\int e^{-i\mathbf{qr}}w_{0}^{2}\left( \mathbf{r}\right) d\mathbf{r\simeq }1
\label{d14}
\end{equation}%
for $\left\vert \mathbf{q}\right\vert \ll l_{c}^{-1},$ where $l_{c}$ is the
localization radius of the function $w_{0}\left( \mathbf{r}\right) $.

The calculation of the sum over $\mathbf{R}_{\mathbf{n}}$ is performed using
the Poisson formula%
\begin{equation}
\sum_{\mathbf{R}_{\mathbf{n}}}e^{i\mathbf{R}_{n}\left( \mathbf{P}_{2}-%
\mathbf{P}_{1}-\mathbf{q}\right) }=\left( \frac{2\pi }{a_{0}}\right)
^{2}\sum_{\mathbf{K}_{\mathbf{n}}}\delta \left( \mathbf{P}_{2}-\mathbf{P}%
_{1}-\mathbf{q-K}_{n}\right) ,  \label{d15}
\end{equation}%
where $\mathbf{K}_{\mathbf{n}}$ is the vector of the reciprocal lattice, $%
\mathbf{K}_{\mathbf{n}}=\frac{2\pi }{a_{0}}\left( n_{x},n_{y}\right) .$The
summation over the vectors of the reciprocal lattice is important because $%
\mathbf{q}$ is not necessarily located in the first Brillouin zone.

As a result, we come to the following expression for the current density (as
it does not depend on time, we omit from now on the variable $\tau $)
\begin{eqnarray}
&&\mathbf{j}_{x,y}\left( \mathbf{q}\right) =e\mathbf{e}_{x,y}\sum_{\mathbf{K}%
_{\mathbf{n}}}\int \delta \left( \mathbf{P}^{\prime }-\mathbf{P}-\mathbf{q-K}%
_{n}\right)  \notag \\
&&\times \left\langle c _{\mathbf{P}}^{\dagger}\left( \tau \right) v_{\mathbf{q}%
}^{x,y}\left( \mathbf{P}\right) c _{\mathbf{P}^{\prime }}\left( \tau
\right) \right\rangle \frac{d\mathbf{PdP}^{\prime }}{\left( 2\pi \right) ^{2}%
},  \label{d16}
\end{eqnarray}%
where $\mathbf{e}_{x,y}=\mathbf{a}_{x,y}/a_{0}$ is the unit vector along $x$
or $y$ bond. In Eq. (\ref{d16}) integration is performed over all $\mathbf{P}
$ and $\mathbf{P}^{\prime }$ inside the first Brillouin zone. The effective
velocity $v_{\mathbf{q}}^{x,y}\left( \mathbf{P}\right) $ equals
\begin{equation}
v_{\mathbf{q}}^{x,y}\left( \mathbf{P}\right) =Ja_{0}\int_{0}^{1}\sin \left[
\left( \mathbf{P+}\alpha \mathbf{q}\right) \mathbf{a}_{x,y}\right] d\alpha .
\label{d17}
\end{equation}%
In the limit $\mathbf{q}\rightarrow 0$, the function $v_{\mathbf{q}%
}^{x,y}\left( \mathbf{p}\right) $ is just the conventional velocity
\begin{equation}
v_{0}^{x,y}\left( \mathbf{P}\right) =Ja_{0}\sin \left( \mathbf{Pa}%
_{x,y}\right) .  \label{d19}
\end{equation}

Averaging in Eq. (\ref{d16}) we reduce the latter to the form
\begin{eqnarray}
&&\mathbf{j}_{x,y}\left( \tau ,\mathbf{q}\right) =2e\mathbf{e}_{x,y}\sum_{%
\mathbf{K}_{n}}\int \delta \left( \mathbf{P}^{\prime }-\mathbf{P}-\mathbf{q-K%
}_{n}\right)  \notag \\
&&\times v_{\mathbf{q}}^{x,y}\left( \mathbf{P}\right) g_{\mathbf{P}^{\prime }%
\mathbf{,P}}\left( 0\right) \frac{d\mathbf{PdP}^{\prime }}{\left( 2\pi
\right) ^{2}},  \label{d21}
\end{eqnarray}%
where
\begin{equation}
g_{\mathbf{P}^{\prime }\mathbf{,P}}\left( 0\right) =-\left\langle c _{%
\mathbf{P}^{\prime }}\left( \tau \right) c _{\mathbf{P}}^{\ast }\left(
\tau \right) \right\rangle .  \label{d22}
\end{equation}

The factor $2$ in Eq. (\ref{d21}) is due to spin.

The main contribution in the integral over $\mathbf{P}$ in Eq. (\ref{d16})
comes from the hot regions and it is again convenient to change to the
variables $c ^{1,2}$, Eq. (\ref{Mod:lagr}) and the momenta $\mathbf{p%
}$ counted from the middle of the edges of the reciprocal lattice. Then, using the symmetry relation
\begin{equation}
g_{\mathbf{p}}^{12}\left( 0\right) =-g_{\mathbf{p}}^{21}\left( 0\right)
\label{d25}
\end{equation}%
and the fact that
\begin{equation}
e^{i\mathbf{Q}_{x}\mathbf{a}_{x,y}}=-e^{i\mathbf{Q}_{y}\mathbf{a}_{x,y}},
\label{d25a}
\end{equation}
Where $\mathbf{Q}_{x}=(\pi/a_0,0),\;\mathbf{Q}_{y}=(0,\pi/a_0)$
Eq. (\ref{d21}) can be written in the form%
\begin{eqnarray}
&&\mathbf{j}_{x,y}\left( \mathbf{q}\right) =2e\mathbf{e}_{x,y}e^{i\mathbf{Q}%
_{x}\mathbf{a}_{x,y}}\sum_{\mathbf{K}_{n}}\int \frac{d\mathbf{p}}{\left(
2\pi \right) ^{2}}\bar{v}_{\mathbf{q}}^{x,y}\left( \mathbf{p}\right) g_{%
\mathbf{p}}^{12}\left( 0\right)  \notag \\
&&\times \Big[\delta \left( \mathbf{q+K}_{n}+\mathbf{Q}_{AF}\right) +\delta
\left( \mathbf{q+K}_{n}-\mathbf{Q}_{AF}\right) \Big],  \label{d26}
\end{eqnarray}%
where
\begin{eqnarray}
&&\bar{v}_{\mathbf{q}}\left( \mathbf{p}\right) =ta_{0}\int_{0}^{1}\sin \left[
\left( \mathbf{p+}\alpha \mathbf{q}\right) \mathbf{a}_{x,y}\right] d\alpha
\notag \\
&=&t\frac{\cos \mathbf{pa}_{x,y}-\cos \left( \left( \mathbf{p+q}\right)
\mathbf{a}_{x,y}\right) }{\mathbf{qe}_{x,y}}.  \label{d27}
\end{eqnarray}

Integrating in Eq. (\ref{d26}) over $\mathbf{p}$ we reduce this equation to
the form%
\begin{eqnarray}
&&\mathbf{j}_{x,y}\left( \mathbf{q}\right) =-8et\overline{D}ie^{i\mathbf{Q}_{x}\mathbf{a%
}_{x,y}}\mathbf{e}_{x,y}  \label{d28} \\
&&\times \frac{\sin ^{2}\left( \mathbf{qa}_{x,y}/2\right) }{\mathbf{qe}_{x,y}%
}\sum_{\mathbf{K}_{n}}\delta \left( \mathbf{q-Q}_{AF}+\mathbf{K}_{n}\right) ,
\notag
\end{eqnarray}%
where
\begin{equation}
\overline{D}=i\int g_{\mathbf{p}}^{12}\left( 0\right) \cos \left( \mathbf{pa}%
_{x,y}\right) \frac{d\mathbf{p}}{\left( 2\pi \right) ^{2}}
\approx\frac{i}{(2\pi\xi)^2}( \langle \hat{c}_1 \hat{c}^\dagger_2 \rangle
-\langle \hat{c}_2 \hat{c}^\dagger_1 \rangle).
  \label{d29}
\end{equation}%
As the order parameter, Eq. (\ref{tm:curord}), is imaginary, the coefficient $\overline{D}$ is
real. The current has peaks at points $\mathbf{Q}_{AF}+\mathbf{K}_{n_{x}}$
and $\mathbf{Q}_{AF}+\mathbf{K}_{n_{y}}$. The function $\mathbf{j}%
_{x,y}\left( \mathbf{q}\right) $ can be further simplified at the peak
values and represented in the form%
\begin{eqnarray}
&&\mathbf{j}_{x,y}\left( \mathbf{q}\right) =-8et\overline{D}ie^{i\mathbf{Q}_{x}\mathbf{a%
}_{x,y}}\mathbf{e}_{x,y}  \label{d29a} \\
&&\times \left( \mathbf{qe}_{x,y}\right) ^{-1}\sum_{\mathbf{K}_{n}}\delta
\left( \mathbf{q-Q}_{AF}+\mathbf{K}_{n}\right) ,  \notag
\end{eqnarray}%
Nevertheless, the sine function can be relevant if the peaks are smeared.

The particle conservation reads
\begin{eqnarray}
&&\mathbf{q\mathbf{j}\left( \tau ,\mathbf{q}\right) }\mathbf{=qj}_{x}\left(
\tau ,\mathbf{q}\right) +\mathbf{qj}_{y}\left( \tau ,\mathbf{q}\right)
\label{d30} \\
&=&8et\overline{D}\sum_{\mathbf{K}_{n}}\left( e^{i\mathbf{Q}_{x}\mathbf{a}_{x}}+e^{i%
\mathbf{Q}_{x}\mathbf{a}_{y}}\right) \delta \left( \mathbf{q-Q}_{AF}+\mathbf{%
K}_{n}\right) =0  \notag
\end{eqnarray}%
The current density can also be written in the real space. It is important
to emphasize that $\mathbf{q}$ is a momentum (not a quasimomentum) and
therefore the current density in the real space $\mathbf{j}\left( \mathbf{r}%
\right) $ can be written for the continuous coordinate $\mathbf{r}$ using
the standard Fourier transform, Eq. (\ref{d13}). A simple calculation leads
to the following expression
\begin{eqnarray}
&&\mathbf{j}_{x,y}\left( \mathbf{r}\right) =-\frac{et\overline{D}a_{0}^{3}e^{i\mathbf{Q}%
_{x}\mathbf{a}_{x,y}}}{\pi ^{2}}\mathbf{e}_{x,y}  \notag \\
&&\times \sum_{\mathbf{R}_{n}}\Big[\cos \left( \mathbf{Q}_{AF}\mathbf{R}%
_{n}\right) \int_{0}^{1}\delta \left( \mathbf{r-R}_{n}+\alpha \mathbf{a}%
_{x,y}\right) d\alpha \Big]  \notag \\
&&  \label{d31}
\end{eqnarray}%
Eq. (\ref{d31}) describes currents circulating around the elementary cells.
The currents oscillate with the period $\mathbf{Q}_{AF}$ and form the bond
current antiferromagnet. This picture corresponds to the one proposed in
Ref. \onlinecite{chakravarty.2001}.

In order to avoid a confusion we would like to note that $\mathbf{j}%
_{x,y}\left( \tau ,\mathbf{r}\right) $ is the two dimensional current
density. The three dimensional current density $\mathbf{J}_{x,y}\left(
\mathbf{r}\right) $ can be written as
\begin{equation}
\mathbf{J}_{x,y}\left( \mathbf{r}\right) =\sum_{m=-\infty }^{\infty }\mathbf{%
j}_{x,y}\left( \mathbf{r,}m\right) \delta \left( z-mc_{0}\right) ,
\label{d32}
\end{equation}%
where $z$ is the coordinate perpendicular to the planes and $c_{0}$ is the
distance between the layers. Provided the currents in different layers are
in phase (the function $\mathbf{j}_{x,y}\left( \mathbf{r,}m\right) $ does
not depend on $m$) one can approximate for rough estimates the 3D current
density as follows%
\begin{equation}
\mathbf{J}_{x,y}\left( \mathbf{r}\right) \simeq \mathbf{j}_{x,y}\left(
\mathbf{r}\right) /c_{0}  \label{d33}
\end{equation}%
In the next subsection we use this approximation in order to visualize
roughly the structure of the magnetic field.

The circulating currents produce magnetic fields that can be measured by
various techniques. As the explicit expression for the spontaneous currents
has been obtained, the magnetic field can be determined without
difficulties. The Fourier transform of the magnetic field $\mathbf{B}\left(
\tau ,\mathbf{q}\right) $ can easily be written using the Maxwell equation%
\begin{equation}
\mathbf{B}\left( \tau ,\mathbf{q}\right) =\frac{4\pi i}{c}\frac{\mathbf{%
q\times J}\left( \tau ,\mathbf{q}\right) }{\mathbf{q}^{2}}  \label{e1}
\end{equation}

With the approximation (\ref{d33}), only the $z$ -component $B^{z}$of the
field,
\begin{equation}
B_{z}\left( \tau ,\mathbf{q}\right) =\frac{4\pi i}{c}\frac{%
q_{x}j_{y}-q_{y}j_{x}}{q^{2}}  \label{e2}
\end{equation}%
is not equal to zero.

\section{Details of calculations for the SF model}
\label{app:sf}
\subsection{Estimate of vertex corrections }
\label{app:sf:ver}
Here we show that the vertex corrections contain a small parameter $\sqrt{[T,\mu,v_s^2/\alpha]/(\alpha/\xi^2)}$ if $\beta\ll\alpha$. As an example we compare the two self-energy diagrams presented in Fig. \ref{Fig:diagrapp} for $\beta=0$.
\begin{figure}[h]
 	\includegraphics[width=0.5\linewidth]{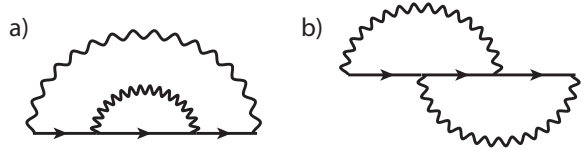}
    \caption{Self-energy diagrams of the same order a) without b) with vertex corrections}
 	\label{Fig:diagrapp}
\end{figure}
The bare propagators for fermions $G_a=(i\varepsilon+\mu-\alpha p_a^2)^{-1}$ are assumed to be much 'sharper' in momentum space than the bosonic ones $\mathfrak{D}=(\omega^2/v_s^2+{\bf p}^2+1/\xi^2)^{-1}$ due to $\alpha/\xi^2\gg\mu$.
The incoming momenta ${\bf p}$ are taken to be $\sim\sqrt{\mu/\alpha}\ll1/\xi$ in magnitude. This allows one to simplify the resulting integrals neglecting the dispersion in the bosonic propagators along $a$, $b$ or both. We get
\begin{gather*}
a)\sim \lambda^4 T^2 \sum_{\varepsilon',\varepsilon''}\int d {\bf p}'d {\bf p}''
\mathfrak{D}(\varepsilon-\varepsilon',{\bf p}-{\bf p}')G_a^2(\varepsilon',{\bf p}')
\mathfrak{D}(\varepsilon'-\varepsilon'',{\bf p}'-{\bf p}'')G_b(\varepsilon'',{\bf p}'')=
\\
\lambda^4 T^2 \sum_{\varepsilon',\varepsilon''}\int d {\bf p}'d {\bf p}''
\frac{1}{(\varepsilon-\varepsilon')^2/v_s^2+({\bf p}-{\bf p}')^2+\frac{1}{\xi^2}}
\frac{1}{(i\varepsilon'+\mu-\alpha(p_a')^2)^2}
\frac{1}{(\varepsilon'-\varepsilon'')^2/v_s^2+({\bf p}'-{\bf p}'')^2+\frac{1}{\xi^2}}
\frac{1}{i\varepsilon''+\mu-\alpha(p_b'')^2}.
\end{gather*}
Using the strong overlap $1/\xi\gg\sqrt{[\mu,T]/\alpha}$ we can neglect ${\bf p}$ and $p'_a$ in the first bosonic propagator and $p'_a$ and $p''_a$ in the second one. We proceed to obtain
\begin{gather*}
\lambda^4 T^2 \sum_{\varepsilon',\varepsilon''}\int d {\bf p}'d {\bf p}''
\frac{1}{(\varepsilon-\varepsilon')^2/v_s^2+(p')_b^2+\frac{1}{\xi^2}}
\frac{1}{(i\varepsilon'+\mu-\alpha(p_a')^2)^2}
\frac{1}{(\varepsilon'-\varepsilon'')^2/v_s^2+(p_a'')^2+(p_b')^2+\frac{1}{\xi^2}}
\frac{1}{i\varepsilon''+\mu-\alpha(p_b'')^2}
\\
\sim\frac{\lambda^4 \frac{v_s^2}{\alpha}}{\Gamma^2} \overline{\sum_{\varepsilon',\varepsilon''}}
\frac{T^2}{\sqrt{(\varepsilon-\varepsilon')^2+\frac{v_s^2}{\xi^2}}}
\frac{1}{(i\varepsilon'+\mu)^{3/2}}
\frac{1}{\sqrt{(\varepsilon'-\varepsilon'')^2+\frac{v_s^2}{\xi^2}}}
\frac{1}{\sqrt{i\varepsilon''+\mu}},
\end{gather*}
where all quantities having dimensions of energy after  $\overline{\sum}$ are normalized to $\Gamma$. For diagram $b)$ we get
\begin{gather*}
b)\sim \lambda^4 T^2 \sum_{\varepsilon',\varepsilon''}\int d {\bf p}'d {\bf p}''
\mathfrak{D}(\varepsilon-\varepsilon',{\bf p}-{\bf p}')G_a(\varepsilon',{\bf p}')G_b(\varepsilon',{\bf p}'')
\mathfrak{D}(\varepsilon'-\varepsilon'',{\bf p}'-{\bf p}'')
G_a(\varepsilon+\varepsilon''-\varepsilon',{\bf p}+{\bf p}''-{\bf p}')\approx
\\
\lambda^4 T^2 \sum_{\varepsilon',\varepsilon''}\int d {\bf p}'d {\bf p}''
\frac{1}{(\varepsilon-\varepsilon')^2/v_s^2+(p')_b^2+\frac{1}{\xi^2}}
\frac{1}{i\varepsilon'+\mu-\alpha(p_a')^2}
\frac{1}{i\varepsilon''+\mu-\alpha(p_b'')^2}
\frac{1}{(\varepsilon'-\varepsilon'')^2/v_s^2+(p_b')^2+\frac{1}{\xi^2}}\cdot
\\
\cdot\frac{1}{i(\varepsilon+\varepsilon''-\varepsilon')+\mu-\alpha(p+p''-p')^2_a}
\\
\sim
\frac{\lambda^4 \left(\frac{v_s^2}{\alpha}\right)^{3/2}}{\Gamma^{5/2}} \overline{\sum_{\varepsilon',\varepsilon''}}
\frac{T^2}{\sqrt{(\varepsilon-\varepsilon')^2+\frac{v_s^2}{\xi^2}}}
\frac{1}{\sqrt{i\varepsilon'+\mu}}
\frac{1}{\sqrt{i\varepsilon''+\mu}}
\frac{1}{(\varepsilon'-\varepsilon'')^2+\frac{v_s^2}{\xi^2}}
\frac{1}{\sqrt{i(\varepsilon+\varepsilon''-\varepsilon')+\mu}}
,
\end{gather*}
where ${\bf p}$ and $p'_a$ are neglected in the first bosonic propagator and $(p'-p'')_a$ and $p''_b$- in the second.
Taking $\Gamma^2 = \lambda^4 \frac{v_s^2}{\alpha}$ we get $1$ in front of the sum for $(a)$ and $\sqrt{(v_s^2/\alpha)/\Gamma}$ for $(b)$. Let us now estimate the Matsubara sums for two cases. For $T\sim\Gamma\gg\mu,v_s/\xi$ sums in $(a)$ and $(b)$ are both of the order $1$ and we get the total result $(b)\sim \sqrt{(v_s^2/\alpha)/\Gamma}\cdot(a)$.

For the calculations in Sec. \ref{sec:sf} a more relevant approximation would be $T\sim\mu,v_s^2/\alpha\ll\Gamma$. It follows then that $v_s/\xi=\sqrt{(v_s^2/\alpha)\cdot(\alpha/\xi^2)}\gg\mu,T,v_s^2/\alpha$. Sums in $(a)$ is estimated as follows. For the one over $\varepsilon'$ one can neglect $\varepsilon'$ in the bosonic propagators. For $\varepsilon''$ the sum evaluated this way diverges, however for an estimate one can use $v_s/\xi/\Gamma$ as a high-frequency cutoff with $\sum T\frac{1}{\sqrt{\varepsilon}}\sim\sqrt{\varepsilon_{max}}$. In total one gets $(a)\sim \left(\left(\frac{v_s}{\xi\Gamma}\right)^{3/2}\sqrt{\frac{T}{\Gamma}}\right)^{-1}$. Estimating $(b)$ in the same way one gets $(b)\sim\sqrt{\frac{v_s^2}{\alpha\Gamma}}\left(\frac{v_s}{\xi\Gamma}\right)^{-5/2}$. Comparing the expressions we obtain $(b)/(a)\sim\sqrt{T/(\alpha/\xi^2)}\ll1$.

For non-zero $\beta$ the fermionic propagators start to disperse along both direction in each region and consequently the argument is not valid. E.g., for $\beta\sim \alpha$ one can ignore the momentum dependence of the first bosonic propagator in $a)$ completely, leading to the same overall form of the answer as in $b)$. On the other hand, if $\beta/\xi^2\ll\mu$ one can ignore $\beta$ in the fermionic propagators, vindicating the argument. Thus the vertex corrections can be neglected at least for $\beta/\xi^2\ll\mu$.

\subsection{Momentum-independent equations for finite $\beta$}
\label{app:sf:eq}
First we calculate the fermionic renormalization related to the self-energy $f^{1(2)}(\varepsilon_n,{\bf p})-\varepsilon_n = i \Sigma^{1(2)}(\varepsilon_n,{\bf p})$.

\begin{gather*}
f^a(\varepsilon_n,{\bf p})-\varepsilon_n = -3 i \lambda^2 T\sum_{\varepsilon_n'}
\int\frac{d {\bf p}'}{(2\pi)^2}
\mathfrak{D}(\varepsilon_n-\varepsilon_n', {\bf p}- {\bf p'}) G^b(\varepsilon_n', {\bf p'})=
\\
-3 i \lambda^2 T\int \frac{d p_a' d p_b'}{(2\pi)^2}
\frac{-1}{\Omega(\varepsilon_n-\varepsilon_n',{\bf p}-{\bf p'})/v_s^2+({\bf p}-{\bf p'})^2+\frac{1}{\xi^2}}
\frac{1}{i f^b(\varepsilon_n',p_a',p_b') - \alpha p_b'^2+\beta p_a'^2+\mu}.
\end{gather*}
Let us first consider the dependence of $f^a$ on $p_a$. For $p_a\sim|\sqrt{(i f +\mu)/\alpha}|\ll|\sqrt{\Omega/v_s^2 +1/\xi^2}|$ one can neglect $p_a$ in the bosonic propagator. It follows then that $f^{a(b)}$ can be taken as independent from $p_{a(b)}$ for such momenta. Consequently, one can then perform the integration over $p_b'$ (as the relevant momenta are $\sim|\sqrt{(i f +\mu)/\alpha}|$ we neglect $p_b'$ in the bosonic propagator):
\[
\int \frac{d p_a'}{4\pi \sqrt{\alpha}}
 \frac{1}{\Omega(\varepsilon_n-\varepsilon_n',p_a',-p_b)/v_s^2+p_a'^2+p_b^2+\frac{1}{\xi^2}}
\frac{i {\rm sgn}[{\rm Re} f^b(\varepsilon_n',p_a')]}
{\sqrt{i f^b(\varepsilon_n',p_a',0) +\beta p_a'^2+\mu}}.
\]
Integral over $p_a'$ has relevant momenta $\sim1/\xi$. We can neglect the dependence of $\Omega$ on $p_a'$ provided that $|\Omega(\varepsilon_n-\varepsilon_n',1/\xi,0)- \Omega(\varepsilon_n-\varepsilon_n',0,0)|/v_s^2\ll1/\xi^2$ (see calculation for $\Omega$ below). For $f$ the condition is $|f(\varepsilon_n',1/\xi,0)- f(\varepsilon_n',0,0)|\ll \beta/\xi^2 + \mu$. Using
\[
\int dx \frac{1}{\sqrt{x^2+a}(x^2+b)} =
\frac{{\rm arctanh}\frac{\sqrt{b-a}}{b}\frac{x}{\sqrt{a+x^2}}}{\sqrt{b}\sqrt{b-a}}
\]
We then obtain:
\begin{gather*}
f^a(\varepsilon_n)-\varepsilon_n =
\frac{3 \lambda^2 T}{2 \pi \sqrt{\alpha \beta}}
\frac{{\rm sgn}[{\rm Re} f^b(\varepsilon_n')]}
{
    \sqrt{\Omega(\varepsilon_n-\varepsilon_n')/v_s^2+\frac{1}{\xi^2}}
    \sqrt{\Omega(\varepsilon_n-\varepsilon_n')/v_s^2+\frac{1}{\xi^2} - (i f^b(\varepsilon_n')+\mu)/\beta}
}
\cdot
\\
\cdot
{\rm arctanh}
\left\{
\sqrt{\frac{
\Omega(\varepsilon_n-\varepsilon_n')/v_s^2+\frac{1}{\xi^2} - (i f^b(\varepsilon_n')+\mu)/\beta
}
{
\Omega(\varepsilon_n-\varepsilon_n')/v_s^2+\frac{1}{\xi^2}
}
}
\right\}
\end{gather*}

Now we turn to calculation of $\Pi(\omega_n,{\bf p})=(\Omega(\omega_n,{\bf p})-\omega_n^2)/v_s^2$:
\begin{gather*}
\Pi(\omega_n,{\bf p}) =
2 \lambda^2 T \sum_{\varepsilon_n,{\bf p}}
G^1(\varepsilon_n+\omega_n, {\bf p}+{\bf q})G^2(\varepsilon_n, {\bf p})
+
G^2(\varepsilon_n+\omega_n, {\bf p}+{\bf q})G^1(\varepsilon_n, {\bf p}).
\end{gather*}
Let us consider the momentum integral in the first term:
\begin{gather*}
\int \frac{d^2 {\bf p}}{(2\pi)^2}
\frac{1}{i f^1(\varepsilon_n+\omega_n) - \alpha (p_1+q_1)^2+\beta (p_2+q_2)^2+\mu}
\frac{1}{i f^2(\varepsilon_n) - \alpha p_2^2+\beta p_1^2+\mu},
\end{gather*}
where we neglect the dependence of $f^{a(b)}$ on the momenta due to arguments presented above. The dependence of the result on ${\bf q}$ is actually controlled by $\beta$: for $\beta\ll\alpha$ one can see that after $p_1\rightarrow p_1-q_1$ the integral has no dependence on ${\bf q}$. We shall take ${\bf q}=0$ in our calculations, which is strictly valid only for small $\beta/\alpha\ll1$. We also introduce a momentum cutoff $\Lambda$ physically motivated by the finite extension of the region, where deviations of the fermionic dispersion from the quadratic form can be ignored. The momentum integral for $\Pi({\bf q},\omega_n)$ is very similar to the DDW susceptibility in the simplified model and we evaluate it in full analogy. First we rewrite the integrand:
\begin{gather*}
\frac{1}
{
(z_+ - \alpha p_1^2 +\beta p_2^2 )
(z - \alpha p_2^2 +\beta p_1^2 )
}
=
\frac{1}
{\alpha z_+ + \beta z + (\beta^2-\alpha^2)p_1^2}
\left[
\frac{\beta}{\beta p_2^2-\alpha p_1^2+z_+}+
\frac{\alpha}{\beta p_1^2-\alpha p_2^2+z}
\right]=
\\
\frac{\alpha}{\alpha+\beta}
\frac{1}
{[(\alpha-\beta) p_1^2-\frac{\alpha z_+ +\beta z}{\alpha+\beta}][\alpha p_2^2 - (z + \beta p_2^2)]}
+
\frac{\alpha}{\alpha+\beta}
\frac{1}
{
    [(\alpha-\beta) p_2^2 -\frac{\alpha z +\beta z_+}{\alpha+\beta}]
    [\alpha p_1^2-(\beta p_1^2+z)]
}
\\
-
\frac{\alpha-\beta}{\alpha+\beta}
\frac{1}
{[(\alpha-\beta) p_1^2-\frac{\alpha z_+ +\beta z}{\alpha+\beta}]
[(\alpha-\beta) p_2^2 -\frac{\alpha z +\beta z_+}{\alpha+\beta}]}
,
\end{gather*}
where $z_+=i f_{\varepsilon+\omega}+\mu,\;z=i f_{\varepsilon+\omega}+\mu$. The result of the integration is:
\begin{gather*}
\lambda^2 T \sum_{\varepsilon}
\frac{4}{\pi^2}
\frac{
    {\rm arctanh}
    \left\{
        \sqrt{\frac{(\alpha^2-\beta^2)\Lambda^2}{i(\alpha f_{\varepsilon+\omega} +\beta f_{\varepsilon})+(\alpha+\beta)\mu}}
    \right\}
    {\rm arctanh}
    \left\{
    \sqrt{
        \frac{i(\alpha f_{\varepsilon+\omega} +\beta f_{\varepsilon})+(\alpha+\beta)\mu}
        {i(\beta f_{\varepsilon+\omega} +\alpha f_{\varepsilon})+(\alpha+\beta)\mu}
    }
        \sqrt{\frac{\alpha \Lambda^2}{\beta \Lambda^2+i  f_{\varepsilon+\omega} + \mu}}
    \right\}
}
{
\sqrt{i(\alpha f_{\varepsilon+\omega} +\beta f_{\varepsilon})+(\alpha+\beta)\mu}
\sqrt{i(\beta f_{\varepsilon+\omega} +\alpha f_{\varepsilon})+(\alpha+\beta)\mu}
}
\\
+\frac{4}{\pi^2}
\frac{
    {\rm arctanh}
    \left\{
        \sqrt{\frac{(\alpha^2-\beta^2)\Lambda^2}{i(\beta f_{\varepsilon+\omega} +\alpha f_{\varepsilon})+(\alpha+\beta)\mu}}
    \right\}
    {\rm arctanh}
    \left\{
    \sqrt{
        \frac{i(\beta f_{\varepsilon+\omega} +\alpha f_{\varepsilon})+(\alpha+\beta)\mu}
        {i(\alpha f_{\varepsilon+\omega} +\beta f_{\varepsilon})+(\alpha+\beta)\mu}
    }
        \sqrt{\frac{\alpha \Lambda^2}{\beta \Lambda^2+i  f_{\varepsilon} + \mu}}
    \right\}
}
{
\sqrt{i(\alpha f_{\varepsilon+\omega} +\beta f_{\varepsilon})+(\alpha+\beta)\mu}
\sqrt{i(\beta f_{\varepsilon+\omega} +\alpha f_{\varepsilon})+(\alpha+\beta)\mu}
}
\\
-\frac{4}{\pi^2}
\frac{
    {\rm arctanh}
    \left\{
        \sqrt{\frac{(\alpha^2-\beta^2)\Lambda^2}{i(\alpha f_{\varepsilon+\omega} +\beta f_{\varepsilon})+(\alpha+\beta)\mu}}
    \right\}
    {\rm arctanh}
    \left\{
        \sqrt{\frac{(\alpha^2-\beta^2)\Lambda^2}{i(\beta f_{\varepsilon+\omega} +\alpha f_{\varepsilon})+(\alpha+\beta)\mu}}
    \right\}
}
{
\sqrt{i(\alpha f_{\varepsilon+\omega} +\beta f_{\varepsilon})+(\alpha+\beta)\mu}
\sqrt{i(\beta f_{\varepsilon+\omega} +\alpha f_{\varepsilon})+(\alpha+\beta)\mu}
}.
\end{gather*}
It can be seen that for $\beta\to0,\;\beta \Lambda^2\to0,\;\alpha \Lambda^2\to \infty$ one recovers our previous result\cite{volkov.2016}. Note that taking the limit $\Lambda\to\infty$ before $\beta\to 0$ leads to a different answer. However, assuming $\Lambda\gtrsim1/\xi$ in the region $\beta\sim\alpha$ we can simplify the answer using $\beta \Lambda^2\gg \mu,T$:
\begin{gather*}
\Pi(\omega_n\neq0,{\bf p})=
\lambda^2 T \sum_{\varepsilon}
\frac{-2i}{\pi}
\frac{
   {\rm sgn}[{\rm Re} (\alpha f_{\varepsilon+\omega} +\beta f_{\varepsilon})]
    {\rm arctanh}
    \left\{
    \sqrt{\frac{\alpha }{\beta}}
    \sqrt{
        \frac{i(\alpha f_{\varepsilon+\omega} +\beta f_{\varepsilon})+(\alpha+\beta)\mu}
        {i(\beta f_{\varepsilon+\omega} +\alpha f_{\varepsilon})+(\alpha+\beta)\mu}
    }
    \right\}
}
{
\sqrt{i(\alpha f_{\varepsilon+\omega} +\beta f_{\varepsilon})+(\alpha+\beta)\mu}
\sqrt{i(\beta f_{\varepsilon+\omega} +\alpha f_{\varepsilon})+(\alpha+\beta)\mu}
}
\\
-\frac{2i}{\pi}
\frac{
   {\rm sgn}[{\rm Re} (\beta f_{\varepsilon+\omega} +\alpha f_{\varepsilon})]
    {\rm arctanh}
    \left\{
    \sqrt{\frac{\alpha }{\beta}}
    \sqrt{
        \frac{i(\beta f_{\varepsilon+\omega} +\alpha f_{\varepsilon})+(\alpha+\beta)\mu}
        {i(\alpha f_{\varepsilon+\omega} +\beta f_{\varepsilon})+(\alpha+\beta)\mu}
    }
    \right\}
}
{
\sqrt{i(\alpha f_{\varepsilon+\omega} +\beta f_{\varepsilon})+(\alpha+\beta)\mu}
\sqrt{i(\beta f_{\varepsilon+\omega} +\alpha f_{\varepsilon})+(\alpha+\beta)\mu}
}
\\
+
\frac{
{\rm sgn}[{\rm Re} (\alpha f_{\varepsilon+\omega} +\beta f_{\varepsilon})]
{\rm sgn}[{\rm Re} (\beta f_{\varepsilon+\omega} +\alpha f_{\varepsilon})]
}
{
\sqrt{i(\alpha f_{\varepsilon+\omega} +\beta f_{\varepsilon})+(\alpha+\beta)\mu}
\sqrt{i(\beta f_{\varepsilon+\omega} +\alpha f_{\varepsilon})+(\alpha+\beta)\mu}
},
\\
\Pi(0,{\bf p})=
\lambda^2 T \sum_{\varepsilon}
\frac{-4i {\rm arctanh}(\sqrt{\beta/\alpha})/\pi{\rm sgn}[{\rm Re}(f_{\varepsilon})]-1 }
{(\alpha+\beta)(i f_{\varepsilon}+\mu)}.
\end{gather*}
As the main objective of current work is to study the effects of finite $\beta$ this expression has been used for numerical calculations. One notes however that $\Pi(0,{\bf p})$ is logarithmically divergent. In what follows we absorb this divergence into the value of $1/\xi^2$ by subtracting $\Pi(0,{\bf p})$ from $\Pi(\omega_n\neq0,{\bf p})$.

Let us now derive the self-consistency equations for the competing order parameters. For the Pomeranchuk instability order parameter $P(\varepsilon_n)$ one has ($P(\varepsilon_n)\equiv P_1(\varepsilon_n)=-P_2(\varepsilon_n)$):
\[
P(\varepsilon_n)=-3 \lambda^2 T\sum_{\varepsilon_n'}\int \frac{d p_a' d p_b'}{(2\pi)^2}
\frac{1}{\Omega(\varepsilon_n-\varepsilon_n')/v_s^2+({\bf p}-{\bf p'})^2+\frac{1}{\xi^2}}
\frac{P(\varepsilon_n')}{(i f(\varepsilon_n') - \alpha p_a'^2+\beta p_b'^2+\mu)^2}
\]
Evaluating the momentum integral under the same assumptions as for the fermion self-energy we obtain:
\begin{gather*}
P(\varepsilon_n)=3 i \lambda^2 T\sum_{\varepsilon_n'}\frac{ P(\varepsilon_n') {\rm sgn}[{\rm Re} f_{\varepsilon'}]}{4\pi\sqrt{\alpha}}
\left\{
\frac{
{\rm arctanh}
    \left\{
        \sqrt{
        \frac{i f_{\varepsilon'}+\mu-\beta(\Omega_{\varepsilon-\varepsilon'}/v_s^2+1/\xi^2)}
        {\beta(\Omega_{\varepsilon-\varepsilon'}/v_s^2+1/\xi^2)}
        }
    \right\}
}
{
    \sqrt{\Omega_{\varepsilon-\varepsilon'}/v_s^2+1/\xi^2}
    (i f_{\varepsilon'}+\mu-\beta(\Omega_{\varepsilon-\varepsilon'}/v_s^2+1/\xi^2))^{3/2}
}-
\right.
\\
\left.
\frac{
\sqrt{\beta}
}
{
    (i f_{\varepsilon'}+\mu)
    (i f_{\varepsilon'}+\mu-\beta(\Omega_{\varepsilon-\varepsilon'}/v_s^2+1/\xi^2))
}
\right\},
\end{gather*}
where we have used:
\[
\int_{-\infty}^{\infty} dx \frac{1}{(x^2+a^2)^{3/2}}\frac{1}{x^2+b^2}=
-2\frac{1}{a^2(a^2-b^2)}
+\frac{2
{\rm arctan}
    \left\{\sqrt{
    \frac{a^2-b^2}{b^2}
    }\right\}
}{b(a^2-b^2)^{3/2}}.
\]
For $\beta\ll\alpha$ we can also obtain the equation for CDW. In this case it is clear that $Q_{CDW}$ is along the diagonal ${\bf Q}_{CDW}=(Q,Q)$ and the equation for the CDW order parameter $C(\varepsilon_n)$ is:
\begin{equation}
\begin{gathered}
C(\varepsilon_n)=-3 \lambda^2 T\sum_{\varepsilon_n'}\int \frac{d p_a' d p_b'}{(2\pi)^2}
\frac{1}{\Omega(\varepsilon_n-\varepsilon_n')+({\bf p}-{\bf p'})^2+\frac{1}{\xi^2}}
\\
\cdot
\frac{C(\varepsilon_n')}
{(i f(\varepsilon_n') - \alpha (p_a'+Q/2)^2+\beta p_b'^2+\mu)
(i f(\varepsilon_n') - \alpha (p_a'-Q/2)^2+\beta p_b'^2+\mu)
}
\end{gathered}
\end{equation}
Evaluating the momentum integral under the same assumptions as for the fermion self-energy we obtain:
\begin{equation}
\begin{gathered}
C(\varepsilon_n)=3 i \lambda^2 T\sum_{\varepsilon_n'}\frac{ C(\varepsilon_n') {\rm sgn}[{\rm Re} f_{\varepsilon'}]}{4\pi\sqrt{\alpha}}
\\
\left\{
\frac{
{\rm arctanh}
    \left\{
        \sqrt{
        \frac{i f_{\varepsilon'}+\mu-\beta(\Omega_{\varepsilon-\varepsilon'}+1/\xi^2)}
        {\beta(\Omega_{\varepsilon-\varepsilon'}+1/\xi^2)}
        }
    \right\}
}
{
    \sqrt{\Omega_{\varepsilon-\varepsilon'}+1/\xi^2}
    (i f_{\varepsilon'}+\mu-\alpha Q^2/4-\beta(\Omega_{\varepsilon-\varepsilon'}+1/\xi^2))
    \sqrt{i f_{\varepsilon'}+\mu-\beta(\Omega_{\varepsilon-\varepsilon'}+1/\xi^2)}
}+
\right.
\\
\left.
\frac{
\sqrt{\beta}
{\rm arctanh}
    \left\{
        \sqrt{
        \frac{\alpha Q^2/4}
        {\alpha Q^2/4 - i f_{\varepsilon'}-\mu}
        }
    \right\}
}
{
    \sqrt{\alpha Q^2/4 - i f_{\varepsilon'}-\mu}
    \sqrt{\alpha Q^2/4}
    (i f_{\varepsilon'}+\mu-\beta(\Omega_{\varepsilon-\varepsilon'}+1/\xi^2))
}
\right\},
\end{gathered}
\end{equation}
where we have used:
\begin{gather*}
\int dx \frac{1}{x^2+c^2}\frac{1}{x^2-a^2}\frac{1}{\sqrt{x^2+b^2}}=
\\
-\frac{
{\rm arctan}
    \left\{
    \frac{\sqrt{b^2-c^2}x}{c\sqrt{b^2+x^2}}
    \right\}
}{c(a^2+c^2)\sqrt{b^2-c^2}}
-\frac{
{\rm arctanh}
    \left\{
    \frac{\sqrt{a^2+b^2}x}{a\sqrt{b^2+x^2}}
    \right\}
}{a(a^2+c^2)\sqrt{a^2+b^2}}.
\end{gather*}

For DDW the equations take form (we use $D(\varepsilon_n)\equiv D^1(\varepsilon_n)=-D^2(\varepsilon_n)$):
\begin{equation}
\begin{gathered}
D(\varepsilon_n)=-3 \lambda^2 T\sum_{\varepsilon_n'}\int \frac{d p_a' d p_b'}{(2\pi)^2}
\frac{1}{\Omega(\varepsilon_n-\varepsilon_n')/v_s^2+({\bf p}-{\bf p'})^2+\frac{1}{\xi^2}}\cdot
\\
\cdot\frac{D(\varepsilon_n')}{(i f(\varepsilon_n') - \alpha p_a'^2+\beta p_b'^2+\mu)(i f(\varepsilon_n') - \alpha p_b'^2+\beta p_a'^2+\mu)}.
\end{gathered}
\end{equation}
First we rewrite the integrand in a similar way to the polarization operator $\Pi({\bf q},\omega_n)$:
\begin{gather*}
\frac{1}
{
(i f_{\varepsilon'} - \alpha p_1^2 +\beta p_2^2 +\mu)
(i f_{\varepsilon'} - \alpha p_2^2 +\beta p_1^2 +\mu)
}
=
\\
\frac{1}
{(\alpha+\beta) i f_{\varepsilon'} +\alpha \mu + \beta \mu + (\beta^2-\alpha^2)p_1^2}
\left[
\frac{\beta}{\beta p_2^2-\alpha p_1^2+i f_{\varepsilon'}+\mu}+
\frac{\alpha}{\beta p_1^2-\alpha p_2^2+i f_{\varepsilon'}+\mu}
\right]=
\\
\frac{\alpha}{\alpha+\beta}
\frac{1}
{[(\alpha-\beta) p_2^2-(if_{\varepsilon'}+\mu)][\alpha p_1^2 - (i f_{\varepsilon'} + \beta p_2^2 +\mu)]}
+
\frac{\alpha}{\alpha+\beta}
\frac{1}
{
    [(\alpha-\beta) p_1^2 -(i f_{\varepsilon'}+ \mu)]
    [\alpha p_2^2-(i f_{\varepsilon'}+\beta p_1^2+\mu)]
}
\\
-
\frac{\alpha-\beta}{\alpha+\beta}
\frac{1}
{[(\alpha-\beta) p_2^2-(if_{\varepsilon'}+\mu)][(\alpha-\beta) p_1^2 -(i f_{\varepsilon'}+ \mu)]}
.
\end{gather*}
If $\alpha-\beta\sim\alpha$ we can neglect the dependence of the bosonic propagator on $p_1(p_2)$ for the first (second) term in the integral and we can neglect the momenta in the bosonic propagator altogether for the third term. We obtain as an intermediate result:
\begin{equation}
\begin{gathered}
D(\varepsilon_n)=\frac{3\lambda^2 T}{\alpha+\beta}\sum_{\varepsilon_n'}
\frac{i D(\varepsilon_n'){\rm sgn}[{\rm Re} f_{\varepsilon'}]}{\pi}
\left\{
\frac{{\rm arctanh}(\sqrt{\beta/\alpha})-i\pi{\rm sgn}[{\rm Re} f_{\varepsilon'}]/2}
{[i  f_{\varepsilon'} +\mu][\Omega_{\varepsilon-\varepsilon'}/v_s^2+1/\xi^2+(i f_{\varepsilon'} +\mu)/(\alpha-\beta)]}
\right.
\\
\left.
+
\frac{\sqrt{\alpha/\beta}}{\alpha-\beta}
\frac{
{\rm arctan}
    \left\{
    \sqrt{
        \frac{(i f_{\varepsilon'} +\mu)/\beta-(\Omega_{\varepsilon-\varepsilon'}/v_s^2+1/\xi^2)}
        {\Omega_{\varepsilon-\varepsilon'}/v_s^2+1/\xi^2}
    }
\right\}
}
{
\sqrt{\Omega_{\varepsilon-\varepsilon'}/v_s^2+1/\xi^2}
[\Omega_{\varepsilon-\varepsilon'}/v_s^2+1/\xi^2+(i f_{\varepsilon'} +\mu)/(\alpha-\beta)]
\sqrt{(i f_{\varepsilon'} +\mu)/\beta-(\Omega_{\varepsilon-\varepsilon'}/v_s^2+1/\xi^2)}
}
\right\}
\\
-
\frac{D(\varepsilon_n')}{4}\frac{1}{i f_{\varepsilon'} +\mu}\frac{1}{\Omega_{\varepsilon-\varepsilon'}/v_s^2+1/\xi^2}
.
\end{gathered}
\end{equation}
We use our assumptions $\alpha-\beta\sim\alpha$ and $|\sqrt{(i f +\mu)/\alpha}|\ll|\sqrt{\Omega +1/\xi^2}|$ to further simplify the answer:
\begin{equation}
\begin{gathered}
D(\varepsilon_n)=\frac{3\lambda^2 T}{\alpha+\beta}\sum_{\varepsilon_n'}
\frac{D(\varepsilon_n')}{\Omega_{\varepsilon-\varepsilon'}/v_s^2+1/\xi^2}
\frac{i{\rm sgn}[{\rm Re} f_{\varepsilon'}]{\rm arctanh}(\sqrt{\beta/\alpha})/\pi+1/4}
{i f_{\varepsilon'} +\mu}.
\end{gathered}
\end{equation}
The result is reminiscent of the expression for $\chi_{DDW}$ in the toy model. To study qualitatively the presence of IDDW within spin-fermion model we use the following equation that can be easily derived assuming $\beta/\xi^2\gg|i f +\mu|$ and incommensurability $(Q,Q)$ along the diagonal using the result (\ref{app:chiddiag}):
\begin{gather*}
D_I(\varepsilon_n)=\frac{3\lambda^2 T}{\alpha+\beta}\sum_{\varepsilon_n'}
\frac{D_I(\varepsilon_n')}{\Omega_{\varepsilon-\varepsilon'}+1/\xi^2}
\left\{
\frac{i{\rm sgn}[{\rm Re} f_{\varepsilon'}]{\rm arctanh}(\sqrt{\beta/\alpha})/\pi+1/4}
{i f_{\varepsilon'} +\mu+\frac{\alpha\beta (\alpha-\beta)Q^2}{(\alpha+\beta)^2}}
\right.
\\
\left.
-\frac{i{\rm sgn}[{\rm Re} f_{\varepsilon'}]/\pi}
{i f_{\varepsilon'} +\mu+\frac{\alpha\beta (\alpha-\beta)Q^2}{(\alpha+\beta)^2}}
\sqrt{
\frac{\frac{\alpha\beta}{(\alpha+\beta)^2}\frac{\alpha\beta Q^2}{\alpha-\beta}}
{i f_{\varepsilon'}+\mu +\frac{\alpha\beta Q^2}{\alpha-\beta}}
}
{\rm arctanh}
\sqrt{
\frac{\frac{\alpha\beta Q^2}{\alpha-\beta}}{if_{\varepsilon'}+\mu +\frac{\alpha\beta Q^2}{\alpha-\beta}}
}
\right\}.
\end{gather*}

Introducing an energy scale $\Gamma=\sqrt{\lambda^2 v_s^2/\alpha}$ (note that in our previous work a larger scale $(\lambda^2 v_s/\sqrt{\alpha})^{2/3}$ has been used) we can bring the equations to a dimensionless form:
\begin{small}
\begin{equation}
\label{app:sf:orders}
\begin{gathered}
P(\varepsilon)=3 i T\sum_{\varepsilon_n'}\frac{P(\varepsilon') {\rm sgn}[{\rm Re} f_{\varepsilon'}]}{4\pi}
\left\{
\frac{
v_s^2/\alpha\cdot
{\rm arctanh}
    \left\{
        \sqrt{
        \frac{(i f_{\varepsilon'}+\mu)v_s^2/\alpha-\beta/\alpha(\Omega_{\varepsilon-\varepsilon'}+a)}
        {\beta/\alpha(\Omega_{\varepsilon-\varepsilon'}+a)}
        }
    \right\}
}
{
    \sqrt{\Omega_{\varepsilon-\varepsilon'}+a}
    [ (i f_{\varepsilon'}+\mu)v_s^2/\alpha- (\Omega_{\varepsilon-\varepsilon'}+a)\beta/\alpha]^{3/2}
}-
\frac{
\sqrt{\beta/\alpha}
}
{
    [i f_{\varepsilon'}+\mu]
    [v_s^2/\alpha (i f_{\varepsilon'}+\mu)-\beta/\alpha (\Omega_{\varepsilon-\varepsilon'}+a)]
}
\right\},
\\
C(\varepsilon)=3 i T\sum_{\varepsilon_n'}\frac{ C(\varepsilon') {\rm sgn}[{\rm Re} f_{\varepsilon'}]}{4\pi}
\\
\left\{
\frac{
v_s^2/\alpha\cdot
{\rm arctanh}
    \left\{
        \sqrt{
        \frac{(i f_{\varepsilon'}+\mu)v_s^2/\alpha-\beta/\alpha(\Omega_{\varepsilon-\varepsilon'}+a)}
        {\beta/\alpha(\Omega_{\varepsilon-\varepsilon'}+a)}
        }
    \right\}
}
{
    \sqrt{\Omega_{\varepsilon-\varepsilon'}+a}
    [(i f_{\varepsilon'}+\mu-\alpha Q^2/4)v_s^2/\alpha-(\Omega_{\varepsilon-\varepsilon'}+a)\beta/\alpha]
    \sqrt{(i f_{\varepsilon'}+\mu)v_s^2/\alpha-(\Omega_{\varepsilon-\varepsilon'}+a)\beta/\alpha}
}+
\right.
\\
\left.
\frac{
\sqrt{\beta/\alpha}
{\rm arctanh}
    \left\{
        \sqrt{
        \frac{\alpha Q^2/4}
        {\alpha Q^2/4 - i f_{\varepsilon'}-\mu}
        }
    \right\}
}
{
    \sqrt{\alpha Q^2/4 - i f_{\varepsilon'}-\mu}
    \sqrt{\alpha Q^2/4}
    [(i f_{\varepsilon'}+\mu)v_s^2/\alpha-(\Omega_{\varepsilon-\varepsilon'}+a)\beta/\alpha]
}
\right\},
\\
D(\varepsilon)=\frac{0.75 T}{1+\beta/\alpha}\sum_{\varepsilon'}
\frac{D(\varepsilon')}{\Omega_{\varepsilon-\varepsilon'}+a}
\frac{4 i{\rm sgn}[{\rm Re} f_{\varepsilon'}]{\rm arctanh}(\sqrt{\beta/\alpha})/\pi+1}
{i f_{\varepsilon'} +\mu},
\\
D_I(\varepsilon)=\frac{0.75 T}{1+\beta/\alpha}\sum_{\varepsilon'}
\frac{D_I(\varepsilon')}{\Omega_{\varepsilon-\varepsilon'}+a}
\left\{
\frac{4 i{\rm sgn}[{\rm Re} f_{\varepsilon'}]{\rm arctanh}(\sqrt{\beta/\alpha})/\pi+1}
{i f_{\varepsilon'} +\mu+\frac{\alpha\beta (\alpha-\beta)Q^2}{(\alpha+\beta)^2}}
\right.
\\
\left.
-\frac{4i{\rm sgn}[{\rm Re} f_{\varepsilon'}]/\pi}
{i f_{\varepsilon'} +\mu+\frac{\alpha\beta (\alpha-\beta)Q^2}{(\alpha+\beta)^2}}
\sqrt{
\frac{\frac{\alpha\beta}{(\alpha+\beta)^2}\frac{\alpha\beta Q^2}{\alpha-\beta}}
{i f_{\varepsilon'}+\mu +\frac{\alpha\beta Q^2}{\alpha-\beta}}
}
{\rm arctanh}
\sqrt{
\frac{\frac{\alpha\beta Q^2}{\alpha-\beta}}{if_{\varepsilon'}+\mu +\frac{\alpha\beta Q^2}{\alpha-\beta}}
}
\right\}.
\end{gathered}
\end{equation}
\end{small}

The equations have been solved numerically by an iteration method with nonlinearities $1/(10|D(\varepsilon')|^2+1),1/(10|P(\varepsilon')|^2+1)$ introduced to r.h.s. of the order parameter equations to enforce convergence below critical temperature. The values of critical temperatures are not affected by this procedure. The number of Matsubara frequencies taken has been $300+1/(\pi T)$, but not larger then $800$.

While solving equations numerically two obstacles were encountered. First, the equations contain nonanalytic functions of a complex argument. To exclude ambiguity, we exclude the half-axis ${\rm Re} z<0,{\rm Im}z=0$ and check that arguments never cross it. In practice this means that the square roots should be evaluated from combinations like $if+\mu$ which always do have an imaginary part and never cross ${\rm Re} z<0,{\rm Im}z=0$ as functions of $\varepsilon$ but not of $(if_1+\mu)*(if_2+\mu)$ or $(if_1+\mu)/(if_2+\mu)$ as these can cross the negative axis as functions of $\varepsilon$.

The second obstacle is that at low $\mu$ spurious solutions for $D$ appear. They don't converge even for large numbers of iterations. However the convergence can be greatly improved by the following trick, which is a simplified version of Newton's method. For $T>T_{Pom,DDW}$ we have an equation:
\[
{\vec X} = A_{ij}{\vec X}_j,
\]
and the Newton's method looks:
\[
{\vec X}_{n+1} ={\vec X}_{n}+ (1-A)^{-1}(A {\vec X}_{n}-{\vec X}_{n}).
\]
Evaluating the matrix $(1-A)^{-1}$ is a rather slow operation. However in our case $A_{ij}\sim(\Omega_{i-j}+a)$ and thus diagonal elements dominate. We can approximately use then:
\[
{\vec X}_{n+1} ={\vec X}_{n}+ diag(1-A_{ii})^{-1}(A {\vec X}_{n}-{\vec X}_{n}).
\]
This allows us to get rid of the spurious solutions and improve convergence to obtain consistent $T_{Pom/DDW}$ values.
\end{widetext}

\end{document}